\documentclass[12pt]{article}
\usepackage{tikz}
\usetikzlibrary{matrix,arrows,decorations.pathmorphing}
\usepackage{jheppub}
\usepackage{amsmath,amssymb,euscript,array,mathrsfs,appendix,ctable}
\usepackage{arydshln}
\usepackage{todonotes}
\usepackage{graphicx}

\usepackage[normalem]{ulem}

\newcommand\VV[1]{\boldsymbol{V}_{\mkern-4mu{#1}}}
\def\mpsu{\mathfrak{psu}}

\def\msl{\mathfrak{sl}}

\def\msl{\mathfrak {sl}}

\def\mh{\mathfrak h}

\def\msu{\mathfrak{su}}
\def\msl{\mathfrak{sl}}

\def\o{\theta}

\def\s{\sigma}

\def\w{\omega}

\newcommand{\IM}{\operatorname{Im}}
\newcommand{\RE}{\operatorname{Re}}
\newcommand{\dn}{\operatorname{dn}}
\newcommand{\sn}{\operatorname{sn}}
\newcommand{\cn}{\operatorname{cn}}

\def\B0{{\boldsymbol 0}}

\def\Dbarslash{\,\,{\raise.15ex\hbox{/}\mkern-12mu {\bar D}}}
\def\Dslash{\,\,{\raise.15ex\hbox{/}\mkern-12mu D}}
\def\delslash{\,\,{\raise.15ex\hbox{/}\mkern-9mu \partial}}
\def\delbarslash{\,\,{\raise.15ex\hbox{/}\mkern-9mu {\bar\partial}}}

\def\ket#1{\left| #1\right\rangle}

\def\II{\mathscr{I}}

\newcommand{\otoprule}{\midrule[\heavyrulewidth]}

\newcommand{\MAT}[1]{\begin{pmatrix} #1\end{pmatrix}}

\newcommand{\EQ}[1]{\begin{equation}\begin{split} #1
\end{split}\end{equation}}

\newcommand{\FIG}[1]{\begin{figure}\begin{center} #1 \end{center}\end{figure}}

\title{${\boldsymbol q}$-Deformation  of the AdS$_{\boldsymbol 5}{\boldsymbol\times}$S$^{\boldsymbol 5}$ Superstring S-matrix and its Relativistic Limit}
\author[a]{Ben Hoare,}
\author[b]{Timothy J. Hollowood}
\author[c]{and J. Luis Miramontes}
\affiliation[a]{Theoretical Physics Group, Blackett Laboratory, Imperial College, London SW7 2AZ, U.K.}
\affiliation[b]{Department of Physics, Swansea University, Swansea, SA2 8PP, U.K.}
\affiliation[c]{Departamento de F\'\i sica de Part\'\i culas and IGFAE,
Universidad
de Santiago de Compostela, 15782 Santiago de Compostela, Spain}

\emailAdd{benjamin.hoare08@imperial.ac.uk} 
\emailAdd{t.hollowood@swansea.ac.uk}
\emailAdd{jluis.miramontes@usc.es}

\abstract{A set of four factorizable non-relativistic  S-matrices for a multiplet of fundamental particles are defined based on the $R$-matrix of the quantum group deformation of the centrally extended superalgebra $\msu(2|2)$. 
The S-matrices are a function of two independent couplings $g$ and $q=e^{i\pi/k}$. The main result is to find the scalar factor, or dressing phase, which ensures that the unitarity and crossing equations are satisfied. 
For generic $(g,k)$, the S-matrices are branched functions on a product of rapidity tori. In the limit $k\to\infty$, one of them is identified with the  
S-matrix describing the magnon excitations on the string world sheet in $\text{AdS}_5\times S^5$, while another is the mirror S-matrix that is needed for the TBA.
In the $g\to\infty$ limit, the rapidity torus degenerates, the branch points disappear and the S-matrices become meromorphic functions, as required by relativistic S-matrix theory. However, it is only the mirror S-matrix which satisfies the correct relativistic crossing equation.
The mirror S-matrix in the relativistic limit is then closely 
related to that of the semi-symmetric space sine-Gordon theory obtained from the string theory by the Pohlmeyer reduction, but has
anti-symmetric rather than symmetric bound states.
The interpolating S-matrix realizes at the quantum level the fact that at the classical level the two theories correspond to different limits of a one-parameter family of symplectic structures of the same integrable system.
}

\notoc
\begin{document}

\begin{flushright}   \small Imperial/TP/11/BH/03
\end{flushright}
\vspace{0.5cm} 

\maketitle

\newpage

\section{Introduction}\label{s1}

\pgfdeclarelayer{background layer} 
\pgfdeclarelayer{foreground layer} 
\pgfsetlayers{background layer,main,foreground layer}

The AdS/CFT correspondence has many facets and its underlying integrability is one of the most remarkable. On the string side this manifests as the integrability of the $1+1$-dimensional QFT on the string world-sheet. The integrability allows one to write down the exact S-matrix for the scattering of magnon degrees-of-freedom: see for example the series of review articles \cite{Beisert:2010jr} and references therein. This S-matrix theory is actually non-relativistic due to the way that the symmetries of the world-sheet theory are fixed. The S-matrix has an underlying Yangian symmetry associated to two copies of the centrally extended Lie superalgebra $\mh=\mpsu(2|2)\ltimes\mathbb R^3$,
just like the S-matrices that describe the (relativistic) principal chiral models in $1+1$-dimensions are associated to the Yangian of a product of two ordinary Lie algebras. The Yangian structure is not easy to describe but one knowns from the example of the principal chiral model that the symmetry structure can be deformed into something more manageable, namely a quantum group $U_q(\mh)$. At the level of the S-matrix the $q$-deformation involves replacing the rational $R$-matrix that describes the magnons by the trigonometric one constructed in \cite{Beisert:2008tw}. This $q$-deformed version of the magnon scattering theory can be viewed as the quantum analogue of the classical fact that the symplectic structure of the integrable system can be continuously deformed. The deformed theory now depends on two coupling constants: $g$ the original coupling of the magnon theory that corresponds to the 't~Hooft coupling of the dual gauge theory and the new quantum deformation parameter $q$. The magnon theory with the Yangian symmetry is then obtained in the rational limit $q\to1$. Another interesting limit is to take $g\to\infty$ keeping $q$ fixed \cite{arXiv:1002.1097}. This was shown in \cite{Hoare:2011fj,Hoare:2011nd} to lead to the quantum S-matrix theory of the relativistic semi-symmetric space sine-Gordon theory. This latter theory is known to be classical equivalent to the string world-sheet theory via the Pohlmeyer reduction but has a different symplectic structure (for work on the Pohlmeyer reduction in this context see \cite{Grigoriev:2007bu,Grigoriev:2008jq,Mikhailov:2007xr,Miramontes:2008wt,Hoare:2009fs,Roiban:2009vh,Hollowood:2009tw,Hollowood:2009sc,Hoare:2010fb,Hollowood:2010rv,Hollowood:2010dt,Hoare:2009rq,Iwashita:2010tg,Iwashita:2011ha,Hollowood:2011fm,Hollowood:2011fq,arXiv:1012.4713,arXiv:1106.4796,Goykhman:2011mq}. The relativistic S-matrix theory was studied in some detail in \cite{Hoare:2011nd} and the S-matrix was shown to fall into a class of S-matrix theories associated to affine quantum groups, in this case for the superalgebra $\msu(2|2)$.

The aim of the present paper is to consider the S-matrix of the interpolating theory, that is for $g$ and $q$ generic, although we take $q$ to be a complex phase. In particular, the scattering of the fundamental states is governed by a product of two of the $R$-matrices of  \cite{Beisert:2008tw} but in order to make a consistent S-matrix this must be multiplied by an overall scalar factor. In the string limit, this is the famous ``dressing phase" which took a lot of work to completely pin down in the magnon theory \cite{Arutyunov:2004vx,Beisert:2005fw,Janik:2006dc,Arutyunov:2006yd,Beisert:2006ib,Beisert:2006qh,Beisert:2006ez,Arutyunov:2006iu,Dorey:2007xn,Kostov:2007kx,Volin:2009uv,Arutyunov:2009kf,Vieira:2010kb,Beisert:2005cw,Hernandez:2006tk,Freyhult:2006vr,Gromov:2007cd,Kruczenski:2009kc}. The main result we present is the dressing phase which incorporates the $q$ deformation. This then determines the S-matrix elements of the interpolating theory for the fundamental states. 
In this discussion it will be important for us to consider both the original magnon theory and the so-called ``mirror theory''. This latter theory arises as the double Wick rotation of the world-sheet theory and is the version of the theory that is relevant for formulating the Thermodynamic Bethe Ansatz \cite{arXiv:0710.1568,Bombardelli:2009ns,arXiv:0902.4458,arXiv:0901.1417,arXiv:0907.2647,arXiv:1012.3995}. The mirror S-matrix is simply the original S-matrix but with a different physical region and therefore a different set of bound-state poles. This turns out to be crucial for our analysis.

The next problem---and one that is not addressed here---will be to find the complete set of bound states of the fundamental states and build up the complete S-matrix by using the bootstrap equations. An important part of this programme has already been completed in a recent paper \cite{deLeeuw:2011jr} which showed that the $R$-matrix for the bound states is determined completely by the quantum group symmetry. In \cite{Hoare:2011nd}, it was conjectured that the S-matrix theory was most naturally defined with $q$ being a root of unity $q=e^{i\pi/k}$, with a positive integer $k$. The spectrum of bound states then consists of a finite set of $k$ atypical, or short, representations of the quantum group. 
The bound-states correspond to symmetric representations in the original S-matrix and anti-symmetric representations in the mirror. 
The infinite set of bound states is then recovered in the $k\to\infty$ limit. In this regard, it could be that the interpolating theory provides a manageable way to regularize the spectrum  of the theory.

This paper is organized as follows. In section 2, we discuss the dispersion relation in the interpolating theory. This naturally describes an elliptic curve whose uniformization leads to the generalized rapidity. We show how the relativistic limit corresponds to a situation where the elliptic curve degenerates to the familiar rapidity cylinder of a relativistic theory. In section 3, we write down the S-matrix for the fundamental states of the interpolating  theory based on the $R$-matrix of \cite{Beisert:2008tw}. In section 4, we consider the unitarity and crossing constraints which are key to pinning down the dressing factor. Particular attention is paid to how the crossing symmetry equation behaves in the relativistic limit. In the interpolating theory there are two possible definitions of crossing symmetry depending on which direction is chosen for the shift by a half period on the rapidity torus. However, with the choice made by the world-sheet theory it is only the mirror version that gives the 
correct crossing equation in the relativistic limit. Section 5 is devoted to the construction of the dressing phase which mirrors very closely what happens in the magnon theory. In general this quantity has a complicated analytic structure since it is a branched function on the rapidity plane. However, we show how the branching disappears in the relativistic limit to give a meromorphic function as expected and required by relativistic S-matrix theory. Section 6 is devoted to a discussion of the bound-state poles of the S-matrix. We show that there is a natural set of vertices that are associated to  bound-states transforming in a set of atypical short representations of the quantum group symmetry.

\section{The Dispersion Relation and Rapidity Torus}\label{s2}

In a relativistic theory, particle states have an energy and momentum that satisfy the familiar dispersion relation $E^2=p^2+m^2$. If we solve this for the energy then we have $E=\pm\sqrt{p^2+m^2}$ meaning that there is a branched double cover of the complex momentum plane. Of course physical particle states have real $p$ and positive energy $E$. For many purposes, particularly for S-matrix theory, it is convenient to find a uniformizing parameterization which removes the branch points. This is achieved by introducing the rapidity $\theta$ with $E=m\cosh\theta$ and $p=m\sinh\theta$. As a complex variable, $\theta$ takes values on the rapidity cylinder $0\leq\IM\theta<2\pi$ and physical values correspond to imposing the reality of $\theta$. In a relativistic factorizable scattering theory, the 2-body S-matrix $S(\theta)$ is a meromorphic function on the infinite un-branched cover of the rapidity cylinder, that is the (relative) rapidity plane $\theta=\theta_1-\theta_2$. The region for which $0\leq\IM\theta\leq\pi$, the {\it physical strip\/}, plays an important role since any poles in this region correspond to bound state processes or anomalous thresholds.

Turning to the non-relativistic S-matrix theory that describes the interpolating theory, the analogues to the dispersion relation and rapidity are a good deal more complicated. In fact we do not know a priori in the interpolating theory what are the energy and momentum of states, all we know is that the fundamental one particle states of the theory are labelled by a pair of variables $\tilde x^\pm$ that satisfy the the non-trivial constraint equation \cite{Beisert:2008tw}
\EQ{
\frac{\tilde x^+}q+\frac q{\tilde x^+}-q\tilde x^--\frac1{q\tilde x^-}+ig(q-q^{-1})\left(\frac{\tilde x^+}{q\tilde x^-}-\frac{q\tilde x^-}{\tilde x^+}\right)=\frac ig\ .
\label{xll}
}
In the above, we should view $g$, a positive real number, and $q$, a complex phase, as two independent couplings.
It is more convenient to work in terms of the shifted and re-scaled variables defined in  \cite{Beisert:2011wq}:
\EQ{
x^\pm=\frac{\tilde g}{g}\tilde x^\pm-\xi\ ,\qquad\xi=-i\tilde g(q-q^{-1})\ ,
\label{nva}
}
where 
\EQ{
\tilde g^2=\frac{g^2}{1-g^2(q-q^{-1})^2}\ .
}
The deformation parameter $q$ will be taken as the complex phase
\EQ{
q=\exp\Big[\frac{i\pi}k\Big]\ ,\qquad k\in\mathbb R>0\ .
}
With the choice above 
both $\tilde g$ and $\xi$ are real numbers with $\tilde g>0$ and $0\leq\xi\leq 1$.
In terms of the new variables, \eqref{xll} becomes the more aesthetic  
\EQ{
q^{-1}\Big(x^++\frac1{x^+}\Big)-q\Big(x^-+\frac1{x^-}\Big)=(q-q^{-1})\Big(\xi+\frac1\xi\Big)\ .
\label{p11}
}
In terms of the representation theory of $U_q(\mh)$, the 3 central charges are determined by $x^\pm$. Following \cite{Beisert:2011wq}, we define the two quantities
\EQ{
U^2=q^{-1}\frac{x^++\xi}{x^-+\xi}=q\frac{\frac1{x^-}+\xi}{\frac1{x^+}+\xi}\ ,\qquad
V^2=q^{-1}\frac{\xi x^++1}{\xi x^-+1}=q\frac{\frac\xi{x^-}+1}{\frac\xi{x^+}+1}\ ,
\label{jww}
}
where the equalities follow from the dispersion relation \eqref{xll}. Then the three central charges are
\EQ{
q^C=V\ ,\qquad P=g\alpha(1-U^2V^2)\ ,\qquad K=g\alpha^{-1}(V^{-2}-U^{-2})\ ,
\label{cch}
}
where $\alpha$ is a constant. These  charges are related via
\EQ{
[C]_q^2-PK=\Big[\frac12\Big]_q^2\ ,
\label{sht}
}
and we have defined $[x]_q=(q^x-q^{-x})/(q-q^{-1})$.
This is precisely the shortening condition for the atypical fundamental 4-dimensional representation of $U_q(\mh)$. 

One advantage of the variables $x^\pm$ is that 
the antipode map (eq.~2.73 of \cite{Beisert:2008tw}) of the quantum group 
which is crucial to the crossing symmetry of the S-matrix is now
particularly simple because it does not depend on $q$:
\EQ{
s(x)=\frac1x\ .
}

Just as in the magnon case, we can solve the constraint \eqref{p11} in terms of a parameterization of $x^\pm$.  
To this end we introduce the map $x(u)$ defined by
\EQ{
x+\frac1x+\xi+\frac1\xi=\frac1\xi\left(\frac{\tilde g}{g}\right)^2 q^{-2iu}\ .
\label{sqr}
}
The pre-factor multiplying $q^{-2iu}$ has been chosen for the convenience of taking particular limits, but can always be absorbed by an additive shift in $u$.
Note that $x(u)$ has square root branch points at $u_\pm$ where
\EQ{
q^{-2iu_\pm}=\left(\frac{g}{\tilde g}\right)^2(\xi\mp1)^2\ ,
}
and $x(u+i0^+)=1/x(u-i0^+)$ along the branch cut which we denote $u\in{\EuScript C}$.
The two branches of $x(u)$ are distinguished by $|x(u)|\gtrless 1$ and the image of a  contour that encircles the branch cut is the unit circle $|x(u)|=1$. 

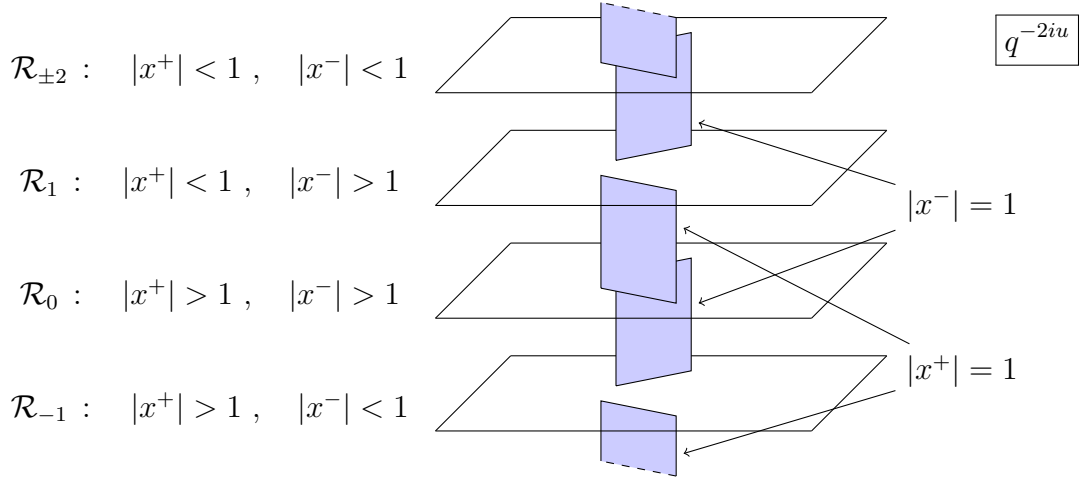
\begin{figure}
\begin{center}
\begin{tikzpicture}[fill=blue!20]
\begin{scope}[yshift=-1.5cm]
\draw [fill]  (3.2,0.9) -- (3.2,1.7) --  (2.2,1.9)  --  (2.2,1.1) ;
\draw [dashed] (3.2,0.9) --  (2.2,1.1) ;
\end{scope}
\draw  (0,0) -- (5,0);
\draw  (5,0) -- (6,1);
\draw  (1,1) -- (6,1);
\draw (1,1) -- (0,0);
\draw [fill] (2.4,0.6) -- (3.4,0.8) -- (3.4,2.3) -- (2.4,2.1) -- (2.4,0.6);
\begin{scope}[yshift=1.5cm]
\draw  (0,0) -- (5,0);
\draw  (5,0) -- (6,1);
\draw  (1,1) -- (6,1);
\draw (1,1) -- (0,0);
\end{scope}
\begin{scope}[yshift=1.5cm]
\draw [fill] (2.2,0.4) -- (3.2,0.2) -- (3.2,1.7) -- (2.2,1.9) -- (2.2,0.4);
\end{scope}
\begin{scope}[yshift=3cm]
\draw  (0,0) -- (5,0);
\draw  (5,0) -- (6,1);
\draw  (1,1) -- (6,1);
\draw (1,1) -- (0,0);
\end{scope}
\begin{scope}[yshift=3cm]
\draw [fill] (2.4,0.6) -- (3.4,0.8) -- (3.4,2.3) -- (2.4,2.1) -- (2.4,0.6);
\end{scope}
\begin{scope}[yshift=4.5cm]
\draw  (0,0) -- (5,0);
\draw  (5,0) -- (6,1);
\draw  (1,1) -- (6,1);
\draw (1,1) -- (0,0);
\end{scope}
\node at (-3,0.3) (i1) {${\cal R}_{-1}\,:\quad |x^+|>1\ ,\quad |x^-|<1$};
\node at (-3,1.8) (i1) {${\cal R}_0\,:\quad |x^+|>1\ ,\quad |x^-|>1$};
\node at (-3,3.3) (i1) {${\cal R}_1\,:\quad |x^+|<1\ ,\quad |x^-|>1$};
\node at (-3,4.8) (i1) {${\cal R}_{\pm2}\,:\quad |x^+|<1\ ,\quad |x^-|<1$};
\node at (8,5.2) (k5) {$\boxed{q^{-2iu}}$};
\node at (7,3) (m1) {$|x^-|=1$};
\node at (7,0.8) (m2) {$|x^+|=1$};
\draw[->] (m1) -- (3.5,1.7);
\draw[->] (m1) -- (3.5,4.1);
\draw[->] (m2) -- (3.3,-0.3);
\draw[->] (m2) -- (3.3,2.7);
\begin{scope}[yshift=4.5cm]
\draw [fill] (2.2,1.2) -- (2.2,0.4) -- (3.2,0.2) -- (3.2,1);
\draw [dashed] (3.2,1) -- (2.2,1.2);
\end{scope}
\end{tikzpicture}
\caption{\small The rapidity torus realized as a 4-fold cover of the cylinder parameterized by $q^{-2iu}$. The top and bottom cut are identified.}
\label{f8} 
\end{center}
\end{figure}

Given the map $x(u)$, we can then solve \eqref{p11} by taking
\EQ{
x^\pm=x\Big(u\pm \frac i2\Big)\ .
}
A function $f(x^+,x^-)$, with $x^+$ and $x^-$ subject to the dispersion relation \eqref{p11}, takes values on a 4-fold cover of the cylinder $u\sim u+ik$ in the complex $u$ plane corresponding to the four possibilities for $|x^\pm|\gtrless1$. Following \cite{Arutyunov:2009kf}, we will denote these sheets as:
\EQ{ 
{\cal R}_{\pm2}\,:\qquad |x^+|<1\ ,\quad |x^-|<1\ , \\
{\cal R}_1\,:\qquad |x^+|<1\ ,\quad |x^-|>1\ , \\
{\cal R}_0\,:\qquad |x^+|>1\ ,\quad |x^-|>1\ , \\
{\cal R}_{-1}\,:\qquad |x^+|>1\ ,\quad |x^-|<1\ , 
}
and it will be useful to think of the label as defined modulo 4 so that ${\cal R}_{-2}\equiv{\cal R}_2$.
This 4-fold covering defines the rapidity torus of the interpolating theory and is illustrated in figure \ref{f8}. The four branches are joined by branch cuts
located along $|x^\pm|=1$. Since the topology is a torus there are non-trivial cycles. For instance if we start on the sheet ${\cal R}_0$, there are two inequivalent ways to reach the sheet ${\cal R}_{\pm2}$. The first involves crossing the cut $|x^+|=1$ to the sheet ${\cal R}_{1}$. Then one crosses the cut $|x^-|=1$ to reach the sheet ${\cal R}_2$. The other path involves crossing the cuts in the other order, first $|x^-|=1$ then $|x^+|=1$ thereby passing through the intermediate sheet ${\cal R}_{-1}$. The distinction between these two paths will become central in the discussion of crossing symmetry.

There are two particular limits of interest:

(i) {\bf The string limit:}  obtained by taking
$q\to1$ or $k\to\infty$. In this limit, $\tilde g\to g$ and $\xi\to 2\pi g/k$ and so $\xi^{-1}q^{-2iu}\to \frac k{2\pi g}+\frac ug$ and so the map $x(u)$ is determined by
\EQ{
x+\frac1x=\frac ug\ .
}
In this case the branch cut ${\EuScript C}$ runs between $u_\pm=\mp2g$.
In this limit, we know that the physical momentum and energy are
\EQ{
p=-i\log\frac{x^+}{x^-}\ ,\qquad E=\frac gi\Big(x^+-\frac1{x^+}-x^-+\frac1{x^-}\Big)
}
and physical values of the parameters, that is $p$ and $E$ real, are obtained when $x^+=(x^-)^*$ which corresponds to $u\in\mathbb R$. A physical particle must have $E>0$ which restricts one to the Riemann sheet ${\cal R}_0$ and if we require positive momentum then $u\in\mathbb R>0$. In this limit, the figure \ref{f8} still applies  but now for the $u$-plane where the cuts corresponding to $|x^\pm|=1$ are parallel. Note that the expressions for the energy and momentum in terms of $x^\pm$ above are modified in the mirror magnon theory \cite{arXiv:0907.2647}.

(ii) {\bf The relativistic limit:}  obtained
by taking $g\to\infty$, in which case $\tilde g\to(2\sin\frac{\pi}k)^{-1}$ and $\xi\to1$. In this limit, the branch points $u_\pm\to\pm\infty$, that is $q^{-2iu_\pm}\to0$ and $\infty$, and 
\EQ{
x(u)\longrightarrow-1 +\frac i{2g\sin\frac\pi k}q^{-iu}+ {\cal O}(g^{-2})\ ,
\label{frq}
}
which is the relativistic parameterization used in the soliton S-matrix of \cite{Hoare:2011nd}, when we identify the relativistic rapidity
as
\EQ{
\theta=\begin{cases} -\frac\pi ku & \text{magnon,}\\ \frac\pi ku & \text{mirror,}\end{cases}
\label{pp2}
}
up to a constant additive shift of $u$ that cancels out from the S-matrix in the relativistic limit since the latter depends only on the difference $\theta_1-\theta_2$. 
The fact that the identification \eqref{pp2} differs in the ordinary and mirror theories by a 
minus sign is, as we shall see, crucial.
In this limit, the branch points $u_\pm$ move out to $\pm\infty$ and the branch cut ${\EuScript C}$ becomes the line $\IM u=0$ modulo $k$.  In the $q^{-2iu}$ plane in figure \ref{f8} the left-hand branch points of the cuts $|x^\pm|=1$ both coalesce at the origin while the right-hand branch points move out to infinity and the covering becomes un-branched.  In this limit, taking the plus sign in \eqref{frq}, the 3 central charges are
\EQ{
C\longrightarrow 0\,, \qquad P\longrightarrow -\alpha\left[\frac{1}{2}\right]_q\, e^{\pm\theta}
\,, \qquad K\longrightarrow  \alpha^{-1} \left[\frac{1}{2}\right]_q\, e^{\mp\theta}\,,
}
with $\pm$ here corresponding to the sign in \eqref{pp2},
so that $P$ and $K$ become identified with the null components of the 2-momentum and the shortening condition \eqref{sht} becomes the familiar relativistic mass-shell condition. 

Another potentially interesting limit is obtained by taking $g\to0$, giving $\tilde g\to g$ and $\xi\to 2g\sin\frac\pi k$. In this case,
\EQ{
x(u)\longrightarrow \frac1{2g\sin\frac\pi k}\Big(q^{-2iu}-1\Big)\ .
}
This  limit is discussed in \cite{Beisert:2011wq} and should also lead to a relativistic  S-matrix although we will not consider it further here.

We have argued that the 4-fold cover of the $u$ cylinder $u\sim u+ik$ on which functions $f(x^+,x^-)$ with $x^\pm$ subject to  \eqref{p11} take values is an elliptic curve. In Appendix \ref{a1}, we show how to uniformize this curve to find the
generalized rapidity $z$, which takes values on the complex plane modulo periodicities in two independent directions:
\EQ{
z\thicksim z+2m\omega_1+2n\omega_2\ ,\qquad m,n\in\mathbb Z\ .
}
In the relativistic limit, after a suitable re-scaling the period $\omega_1$ diverges and $\omega_2$ remains constant which manifests the degeneration of the rapidity torus to the relativistic rapidity cylinder. 
We remark that the rapidity $z$ itself is not such a useful coordinate in the non-relativistic interpolating S-matrix theory because the 2-body S-matrix is not simply a meromorphic function on a product of two rapidity $z$-planes. In fact, the S-matrix turns out to take values on a branched cover of the rapidity plane which defines an infinite genus Riemann surface. So although $z$ uniformizes the dispersion relation itself it does not uniformize the S-matrix. Nevertheless we find it useful to write the rapidity dependence of quantities in terms of $z$ where appropriate. We note that the antipode operation involves a shift by a half period:
\EQ{
s(x^\pm(z))=x^\pm(z+\varepsilon\omega_2)=\frac1{x^\pm(z)}\ ,\qquad\varepsilon=\pm1\ ;
\label{apo}
}
in particular, $s:\,{\cal R}_{n}\to{\cal R}_{n\pm2}$,
a fact that will be important for the discussion of crossing symmetry.

\section{The Interpolating S-matrix}

The S-matrix is constructed using a product of two copies of the fundamental  $R$-matrix of the quantum group deformation of the triply extended superalgebra $\mh=\mpsu(2|2)\ltimes{\mathbb R}^3$ in \cite{Beisert:2008tw}, with the central extensions identified. Each particle multiplet is $4\times 4$ dimensional, and carries a product of two of the four-dimensional fundamental representations of the algebra, denoted $\langle0,0\rangle$ in \cite{Beisert:2008tw}.

For completeness we now write down the fundamental $R$-matrix in terms of the new coordinates $x^\pm$.\footnote{These are related to $x^\pm$ in the reference \cite{Beisert:2008tw}, which earlier we denoted as $\tilde x^\pm$, as in \eqref{nva}.} The four states, two bosonic and two fermionic, are denoted $\{|\phi^a\rangle,|\psi^a\rangle\}$, $a=1,2$. 
Here, we write down the related matrix $\check R=P\cdot R$ where $P$ is the graded permutation matrix
\EQ{
\check R(z_1,z_2):\ \VV{1}(z_1)\otimes \VV{1}(z_2)\longrightarrow \VV{1}(z_2)\otimes \VV{1}(z_1)\ ,
\label{ppx}
}
where $\VV{1}(z_1)$ is the 4-dimensional module spanned by the basis $\{|\phi^a\rangle,|\psi^a\rangle\}$. We then have
\EQ{ 
\check R\ket{\phi^a\phi^a}&=A\ket{\phi^a\phi^a}\ ,\qquad \check R \ket{\psi^\alpha\psi^\alpha}=D\ket{\psi^\alpha\psi^\alpha}\ ,\\
\check R \ket{\phi^1\phi^2}&=\frac{q(A-B)}{q^2+1}\ket{\phi^2\phi^1}+
\frac{q^2 A+B}{q^2+1}\ket{\phi^1\phi^2}+\frac{C}{1+q^2}\ket{\psi^1\psi^2}-\frac{qC}{1+q^2}\ket{\psi^2\psi^1}\ ,\\
\check R \ket{\phi^2\phi^1}&=\frac{q(A-B)}{q^2+1}\ket{\phi^1\phi^2}
+\frac{q^2 B+A}{q^2+1}\ket{\phi^2\phi^1}-\frac{qC}{1+q^2}\ket{\psi^1\psi^2}+\frac{q^2C}{1+q^2}\ket{\psi^2\psi^1}\ ,\\
\check R \ket{\psi^1\psi^2}&=\frac{q(D-E)}{q^2+1}\ket{\psi^2\psi^1}
+\frac{q^2 D+E}{q^2+1}\ket{\psi^1\psi^2}+\frac{F}{1+q^2}\ket{\phi^1\phi^2}-\frac{qF}{1+q^2}\ket{\phi^2\phi^1}\ ,\\
\check R \ket{\psi^2\psi^1}&=\frac{q(D-E)}{q^2+1}\ket{\psi^1\psi^2}
+\frac{q^2 E+D}{q^2+1}\ket{\psi^2\psi^1}-\frac{qF}{1+q^2}\ket{\phi^1\phi^2}+\frac{q^2F}{1+q^2}\ket{\phi^2\phi^1}\ ,\\
\check R \ket{\phi^a\psi^\alpha}&=G\ket{\psi^\alpha\phi^a}+H\ket{\phi^a\psi^\alpha}\ ,\qquad
\check R \ket{\psi^\alpha\phi^a}=K\ket{\psi^\alpha\phi^a}+L\ket{\phi^a\psi^\alpha}\ .
\label{jjs}
}
The functions $A$ to $L$ are defined in  \cite{Beisert:2008tw} as $A_{12}$ to $L_{12}$   (with $R_{12}^0=1$) with respect to the original parameters $\tilde x_i^\pm$. In terms of the new parameters $x_i^\pm$, we have\footnote{Note that in terms of the coordinates $x^\pm$ the quantity $\gamma$ defined in  \cite{Beisert:2008tw} is $\sqrt{-i\alpha q^{1/2}UV(x^+-x^-)}$ where $\alpha$ is a constant that cancels out of the S-matrix.}
\EQ{
A&=\frac{U_1V_1}{U_2V_2}\cdot\frac{x_2^+-x_1^-}{x_2^--x_1^+}\ ,\quad
B=A\Big(1-(1+q^{-2})\cdot\frac{x_2^+-x_1^+}{x_2^+-x_1^-}\cdot\frac{x_2^--\frac1{x_1^+}}{x_2^--\frac1{x_1^-}}\Big)\ ,\\
C&=F=i(1+q^{-2})\Big(\frac{U_1V_1}{U_2V_2}\Big)^{3/2}\cdot\frac{1-\frac{x_2^+}{x_1^+}}{x_2^--\frac1{x_1^-}}\cdot\frac{\sqrt{(x_1^+-x_1^-)(x_2^+-x_2^-)}}{x_2^--x_1^+}\  ,\\  D&=-1\ ,\quad E=-1+\frac{1+q^{-2}}{U_2^2V_2^2}\cdot\frac{x_2^+-x_1^+}{x_2^--x_1^+}\cdot
\frac{x_2^+-\frac1{x_1^-}}{x_2^--\frac1{x_1^-}}\ , \\
G&=q^{-1/2}\frac1{U_2V_2}\cdot\frac{x_2^+-x_1^+}{x_2^--x_1^+}\ ,\quad H=K=\sqrt{\frac{U_1V_1}{U_2V_2}}\cdot\frac{\sqrt{(x_1^+-x_1^-)(x_2^+-x_2^-)}}{x_2^--x_1^+}\ ,\\  L&=q^{1/2}U_1V_1\cdot\frac{x_2^--x_1^-}{x_2^--x_1^+}\ .
\label{eqnsb}}

The S-matrix takes the form expected for a theory with a product $U_q(\mh)\times U_q(\mh)$ symmetry\footnote{Similar to the principal chiral model which has a $G\times G$ symmetry.}
\EQ{
S(z_1,z_2)=\frac{U_1^2}{U_2^2}\cdot\frac{Z(z_1,z_2)}{\sigma(z_1,z_2)^2}\cdot\check R(z_1,z_2)\,\otimes_\text{gr}\,\check R(z_1,z_2)\ ,
\label{pcm}
}
where $Z(z_1,z_2)$ is a scalar factor which compensates for the fact that 
the product of two $\check R(z_1,z_2)$ matrices would have a double pole at $x_1^+=x_2^-$ and this should be a simple pole since it corresponds to the $s$-channel bound state transforming in the short, or atypical, symmetric representation $\langle 1,0\rangle$ of $U_q(\mh)$:\footnote{When we make a statement like this we always mean a product of two of these representations, one for each $U_q(\mh)$ factor.}
\EQ{
Z(z_1,z_2)=\frac{x_2^--x_1^+}{x_2^+-x_1^-}\cdot\frac{x_1^-x_2^+-1}{x_1^+x_2^--1}\ .
\label{fzz}
}
In \eqref{pcm}, the tensor product respects the fermionic grading of the states. The multiplication by the factors involving $U_i$ in \eqref{jww},
\EQ{
U_i^2=q^{-1}\frac{x_i^++\xi}{x_i^-+\xi}=q\frac{\frac1{x_i^-}+\xi}{\frac1{x_i^+}+\xi}\ ,
}
is needed so that the dressing factor $\sigma(z_1,z_2)$ agrees with the definition  in \cite{Volin:2009uv,Vieira:2010kb,Arutyunov:2009kf} when the magnon limit is taken.

Notice that once multiplied by $Z(z_1,z_2)$ the product of two $\check R(z_1,z_2)$ matrices also has a simple pole at $x_1^-=x_2^+$. This is not on the physical sheet for the magnon S-matrix but becomes the $s$-channel bound-state pole for the mirror S-matrix. It corresponds to the bound state transforming in the short, or atypical, anti-symmetric representation $\langle 0,1\rangle$ of $U_q(\mh)$. It is important to note that  the S-matrix of the mirror theory is just an analytic continuation of the S-matrix of the magnon theory, the difference lies in the definition of the physical values of rapidity and the physical sheet and consequently the bound states and bootstrap are different.

To fix our conventions, we write down the S-matrix in the so-called $\msu(2)$ and $\msl(2)$ sectors in the string limit.\footnote{Since there are many different conventions in the literature---$\sigma$ can be in the denominator or numerator and $x_1$ can be swapped with $x_2$---appendix \ref{app2} provides a discussion of how our conventions relate to other relevant works.} These algebras are the bosonic subalgebras of (complexified) $\msu(2|2)$ under which $\phi^a$ transform as $(1,0)$ and $\psi^a$ as $(0,1)$.\footnote{The representations are labelled with $(2j_1,2j_2)$.}
The S-matrix element for the ``$\msu(2)$ sector" corresponds to the particular element
$|\phi^a\phi^a;z_1\rangle\otimes\ket{\phi^a\phi^a;z_2}\to|\phi^a\phi^a;z_2\rangle\otimes\ket{\phi^a\phi^a;z_1}$:
\EQ{
S_{\msu(2)}(z_1,z_2)&=\frac{U_1^2}{U_2^2}\cdot\frac{Z(z_1,z_2)}{\sigma(z_1,z_2)^2}\cdot A(z_1,z_2)^2\\
&=\frac1{\sigma(z_1,z_2)^2}\cdot\frac{x_1^+x_2^-}{x_1^-x_2^+}\cdot\frac{x_1^--x_2^+}{x_1^+-x_2^-}\cdot
\frac{1-\frac1{x_1^-x_2^+}}{1-\frac1{x_1^+x_2^-}}\\ &=\frac1{\sigma(z_1,z_2)^2}\cdot\frac{x_1^+x_2^-}{x_1^-x_2^+}\cdot\frac{u_1-u_2-i}{u_1-u_2+i}\ .
\label{xpp}
}
This form manifests the $s$-channel pole at $x_1^+=x_2^-$, or $u_1-u_2=-i$ on the sheet ${\cal R}_{0,0}$. The 
fact that this pole appears in the symmetric channel of the (two copies of) $\msu(2)\subset\msu(2|2)$,  identifies the bound state as the atypical symmetric $\langle1,0\rangle$ representation of $\msu(2|2)$. 
On the other hand, for the ``$\msl(2)$ sector" corresponding to the particular element
$|\psi^a\psi^a;z_1\rangle\otimes\ket{\psi^a\psi^a;z_2}\to|\psi^a\psi^a;z_2\rangle\otimes\ket{\psi^a\psi^a;z_1}$, we have
\EQ{
S_{\msl(2)}(z_1,z_2)&=\frac{U_1^2}{U_2^2}\cdot\frac{Z(z_1,z_2)}{\sigma(z_1,z_2)^2}\cdot D(z_1,z_2)^2\\
&=\left(\frac{1-\frac1{x_1^+x_2^-}}{1-\frac1{x_1^-x_2^+}}\sigma(z_1,z_2)\right)^{-2}\cdot\frac{x_1^+-x_2^-}{x_1^--x_2^+}\cdot
\frac{1-\frac1{x_1^+x_2^-}}{1-\frac1{x_1^-x_2^+}}\\ &=\left(\frac{1-\frac1{x_1^+x_2^-}}{1-\frac1{x_1^-x_2^+}}\sigma(z_1,z_2)\right)^{-2}\cdot\frac{u_1-u_2+i}{u_1-u_2-i}\ ,
\label{xpp2}
}
matching eq.~1.2 and 1.4 of \cite{arXiv:0901.1417} (with a re-scaling of $u$).
This form manifests the mirror $s$-channel pole at $x_1^-=x_2^+$, or $u_1-u_2=i$ which lies off the physical sheet for the magnon theory. 
Note that in this case in the bosonic sector for each $\msu(2|2)$ factor the bound state couples to 
the anti-symmetric channel $\ket{\phi^a}\otimes|\phi^b\rangle$. For instance, the $\msu(2)$ element in \eqref{xpp} involving the symmetric combination  $\ket{\phi^a}\otimes\ket{\phi^a}$ does not have a pole there.
The bound-state representation is then the atypical anti-symmetric $\langle0,1\rangle$ representation of $\msu(2|2)$.

\section{Crossing and Unitarity}

In order to have a consistent S-matrix, we need to determine the scalar factor---the ``dressing phase"---denoted $\sigma(z_1,z_2)$ in the last section. It is important to recognise that the $z_i$ dependence of $\sigma(z_1,z_2)$ does not mean that it is valued on the rapidity torus. In fact the dressing phase takes values on an infinite branched cover of the rapidity torus. 
In a relativistic scattering theory things are simpler and the covering becomes un-branched. In this case, $\sigma(\theta_1,\theta_2)\equiv\sigma(\theta)$, with $\theta\equiv\theta_1-\theta_2$, is a meromorphic infinite cover of the rapidity cylinder identified with the complex $\theta$ plane.

We define S-matrix elements, so that
\EQ{
S_{ij}^{kl}(z_1,z_2):\ |i,z_1\rangle\otimes|j,z_2\rangle\longrightarrow
|k,z_2\rangle\otimes|l,z_1\rangle\ .
}
Here, $i,j,\ldots=1,2,\ldots,16$ label states of the $16$-dimensional representation formed from the product of fundamental representations $\VV{1}$ for each of the two symmetry algebras. The unitarity condition requires that\footnote{The following condition is commonly called the ``unitarity condition". However, it does not say that the S-matrix is a unitary S-matrix and is perhaps better termed ``braiding unitarity" or ``$R$-matrix unitarity" in the context of quantum groups. The S-matrix axiom of Hermitian Analyticity usually guarantees the equivalence between braiding and matrix unitarity. It is known that S-matrices built from quantum group intertwiners when $q$ is a complex phase may not satisfy Hermitian Analyticity, and na\"\i vely the present $R$ matrix is no  different, as pointed out in \cite{Beisert:2008tw}. However, Hermitian Analyticity only need hold in some particular basis in the one-particle Hilbert space \cite{Miramontes:1999gd}, and this basis freedom has not been exploited in the present case to give a definitive answer to the question of unitarity.}
\EQ{
S_{ij}^{mn}(z_1,z_2)S_{mn}^{kl}(z_2,z_1)=\delta_i^k\delta_j^l\ .
\label{run}
}

Now we turn to the crucial implications of crossing symmetry. The action of charge conjugation on particles involves taking the antipode operation on the parameters \eqref{apo}, $s:\ x^\pm\to1/x^\pm$. However, this operation requires care because we should be mindful of the path followed on the rapidity torus from $x^\pm$ to the antipode point. This becomes clearer if we think of the uniformized variable $z$. The antipode point is at $z+\omega_2$, however, this can be reached in more than one way since on the torus $z+\omega_2\equiv z-\omega_2$. Although the $R$-matrix and factor $Z$ in \eqref{fzz} are periodic on the torus, the S-matrix itself is {\it not\/}, due to the dressing phase $\sigma(z_1,z_2)$, and so one should be careful to distinguish the points $z\pm\omega_2$. 
We will investigate this point in more detail below.

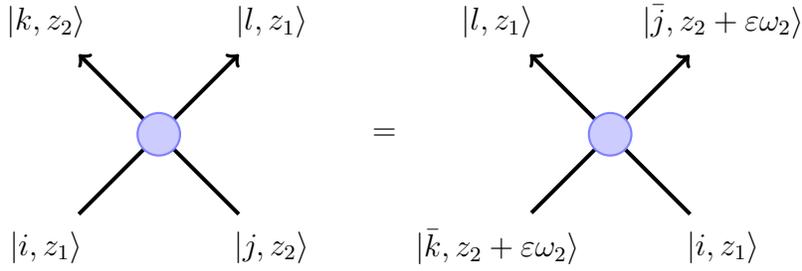
\begin{figure}[t]
\begin{center}
\begin{tikzpicture} [line width=1.5pt,inner sep=2mm,
place/.style={circle,draw=blue!50,fill=blue!20,thick}]
\begin{pgfonlayer}{foreground layer}
\node at (1.5,1.5) [place] (sm) {}; 
\end{pgfonlayer}
\node at (0,0) (i1) {$|i,z_1\rangle$};
\node at (3,0) (i2) {$|j,z_2\rangle$};
\node at (0,3) (i3) {$|k,z_2\rangle$};
\node at (3,3) (i4) {$|l,z_1\rangle$};
\draw[->] (i1) -- (i4);
\draw[->] (i2) -- (i3);
\node at (4.5,1.5) (k1) {$=$};
\begin{pgfonlayer}{foreground layer}
\node at (7.5,1.5) [place] (sm) {}; 
\end{pgfonlayer}
\node at (6,0) (j1) {$|\bar k,z_2+\varepsilon\omega_2\rangle$};
\node at (9,0) (j2) {$|i,z_1\rangle$};
\node at (6,3) (j3) {$|l,z_1\rangle$};
\node at (9,3) (j4) {$|\bar j,z_2+\varepsilon\omega_2\rangle$};
\draw[->] (j1) -- (j4);
\draw[->] (j2) -- (j3);\end{tikzpicture}
\caption{\small Crossing symmetry of the basic 2-body S-matrix. Note that the right-hand diagram is just a $\frac\pi2$ rotation of the left-hand diagram.}
\label{f3} 
\end{center}
\end{figure}

To start with, here is one way to write the crossing relation in a relativistic theory:
\EQ{
S_{ij}^{kl}(\theta_1,\theta_2)=S_{\bar ki}^{l\bar j}(\theta_2+i\pi,\theta_1)\ ,
\label{rcr1}
}
where the charge conjugate states are $|\bar i,\theta\rangle={\cal C}_{\bar ii}|i,\theta\rangle$ with ${\cal C}$ the charge conjugation matrix, and 
$S(\theta_1,\theta_2)\equiv S(\theta_1-\theta_2)$ due to relativistic invariance. This gives the familiar form
\EQ{
S_{ij}^{kl}(\theta)=S_{\bar ki}^{l\bar j}(i\pi-\theta)\ .
\label{fff}
}
So here the analogue of the antipode operation involves $\theta\to \theta+i\pi$ for the second particle.
Of course, as points on the rapidity cylinder, $\theta\pm i\pi$ are equivalent, but the S-matrix is not periodic on the cylinder and so the choice matters.
The implication is that 
when we consider the interpolating S-matrix the crossing relation will involve taking $x^\pm\to1/x^\pm$ but implicitly this entails choosing a direction on the rapidity torus, either $z\to z+\omega_2$ or $z\to z-\omega_2$. If we start on the Riemann sheet with $|x^\pm|>1$, that is ${\cal R}_0$, then crossing symmetry involves changing to the sheet with $|x^\pm|<1$, that is ${\cal R}_{\pm2}$.
As we explained in section \ref{s2} the two directions on the torus correspond to crossing the curves $|x^\pm|=1$ in a particular order, which can also be thought of as the choice of which intermediate sheet to cross. The relation with the uniformized variable is
\EQ{
z\to z\pm\omega_2:\qquad{\cal R}_0 \longrightarrow\ {\cal R}_{\pm1}\ \longrightarrow\ {\cal R}_{\pm2}\ .
\label{chs}
}
In addition, an important point is that for the rapidity torus itself ${\cal R}_2$ is identified with ${\cal R}_{-2}$ but for the S-matrix this identification is no longer valid and the sheets ${\cal R}_{\pm2}$ become distinct. In fact, as we shall see, by analytic continuation and
by repeated shifts by $\omega_2$ we move out onto an infinite set of sheets ${\cal R}_n$ for any integer $n$. Note that the interpolating S-matrix, being a function of $z_1$ and $z_2$, is valued on a product of the sheets which is denoted ${\cal R}_{m,n}$.
An important subtlety that does not arise in the relativistic limit is that the S-matrix will turn out to have branch points and associated cuts on some of the sheets. 
This means that when we write the crossing equations there is an implicit choice for the path of the analytic continuation. 
It is not immediately obvious what sign in \eqref{chs} is needed to match \eqref{rcr1}, so for the moment, we will write 
$z_2\to z_2+\varepsilon\omega_2$, with $\varepsilon=\pm1$, and then the analogue of \eqref{rcr1} is\footnote{Note that we could equivalently write the crossing equation in the form
$S_{ij}^{kl}(\theta_1,\theta_2)=S_{j\bar l}^{\bar ik}(\theta_2,\theta_1-i\pi)$ which becomes
$S_{ij}^{kl}(z_1,z_2)=S_{j\bar l}^{\bar ik}(z_2,z_1-\varepsilon\omega_2)$ instead. This other way of writing the crossing equation is equivalent to
\eqref{csm} and leads to the same constraint on the dressing phase below in \eqref{css2}.}
\EQ{
S_{ij}^{kl}(z_1,z_2)=S_{\bar ki}^{l\bar j}(z_2+\varepsilon\omega_2,z_1)\ ,
\label{csm}
}
which is illustrated in figure~\ref{f3}. 

The $\check R$-matrix itself satisfies the unitarity condition
\EQ{
\check R_{ij}^{mn}(z_1,z_2)\check R_{mn}^{kl}(z_2,z_1)=\delta_i^k\delta_j^l
\label{run2}
}
and the crossing relation
\EQ{
\check R_{ij}^{kl}(z_1,z_2)=p(z_1,z_2)
\check R_{\bar ki}^{l\bar j}(z_2\pm\omega_2,z_1)\ .
\label{rcr}
}
The choice of sign here does not matter since the $R$-matrix is an honest function on the rapidity torus. The
charge conjugation matrix is
\EQ{
{\cal C}=\MAT{0&q^{1/2}&0&0\\ -q^{-1/2}&0&0&0\\ 0&0&0&-q^{1/2}\\ 0&0&q^{-1/2}&0}
}
and we have defined a function
\EQ{
p(z_1,z_2)=q^{-1}\cdot\frac{x_1^+-x_2^+}{x_1^+-x_2^-}\cdot\frac{1-\frac1{x_1^-x_2^+}}{1-\frac1{x_1^-x_2^-}}\ .
}

The factor $Z(z_1,z_2)$ in \eqref{pcm} is also an honest function on the rapidity torus and is completely inert with regard to unitarity and crossing, meaning
\EQ{
Z(z_1,z_2)Z(z_2,z_1)=1\ ,\qquad Z(z_1,z_2)=Z(z_2\pm\omega_2,z_1)\ .
}
Putting all the unitarity and crossing relations together means that the  dressing factor must satisfy the unitarity condition
\EQ{
\sigma(z_1,z_2)\sigma(z_2,z_1)=1
\label{unit1}
}
and the crossing relation
\EQ{
\sigma(z_1,z_2+\varepsilon\omega_2)=U_1^{2}p(z_1,z_2)\sigma(z_2,z_1)\ .
\label{cs22}
}
If we use unitarity and swap $z_1$ and $z_2$, we can write
this as
\EQ{
\sigma(z_1+\varepsilon\omega_2,z_2)\sigma(z_1,z_2)=\frac1{U_2^{2}p(z_2,z_1)}
=\frac{x_2^-+\xi}{x_2^++\xi}\cdot \frac{x_1^--x_2^+}{x_1^--x_2^-}\cdot\frac{1-\frac1{x_1^+x_2^+}}{1-\frac1{x_1^+x_2^-}}\ ,
\label{css2}
}
a form that is favoured in the string theory literature. Above, we have employed the identity
\EQ{
q^{-1}(x_1^+-x_2^+)\Big(1-\frac1{x_1^+x_2^+}\Big)=q(x_1^--x_2^-)\Big(1-\frac1{x_1^-x_2^-}\Big)\ ,
\label{trick}
}
which follows from \eqref{p11}. In appendix \ref{app2}, we show that \eqref{css2} with $\varepsilon=+1$ is equivalent to the crossing equation written by Janik \cite{Janik:2006dc} and in other subsequent works.
In particular, in the string limit, $\xi\to0$, the crossing equation \eqref{css2} is precisely the one written in both \cite{Arutyunov:2009kf}, eq.~2.8, and \cite{Vieira:2010kb}, eq.~3.4.

\vspace{0.5cm}
\noindent{\bf The crossing ambiguity $\boldsymbol\varepsilon$}

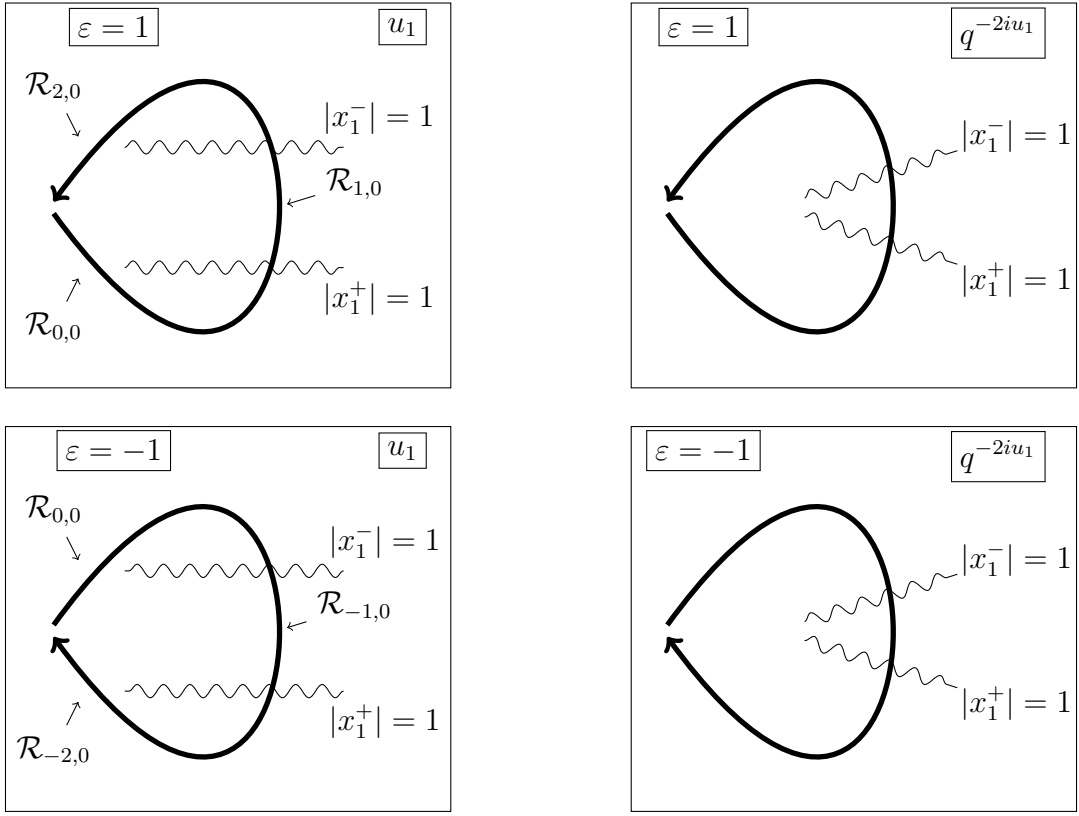
\begin{figure}[t]
  \begin{minipage}[t]{.47\textwidth}
    \begin{center}
\begin{tikzpicture}[scale=0.8]
\clip (-4,-3.5) rectangle (3.8,3.5);
\draw (-3.8,-3) rectangle (3.6,3.4);
\node at (-2,3) (k1) {$\boxed{\varepsilon=1}$};
\node at (2.8,3) (k5) {$\boxed{u_1}$};
\node at (2.4,1.5) (m1) {$|x_1^-|=1$};
\node at (2.4,-1.5) (m1) {$|x_1^+|=1$};
\node at (-3,-2) (k2) {${\cal R}_{0,0}$};
\node at (-2.5,-1) (n2) {};
\draw [->] (k2) -- (n2);
\node at (-3,2) (k3) {${\cal R}_{2,0}$};
\node at (-2.5,1) (n3) {};
\draw [->] (k3) -- (n3);
\node at (2,0.4) (k4) {${\cal R}_{1,0}$};
\node at (0.7,0) (n4) {};
\draw [->] (k4) -- (n4);
\node at (-2,1) (j1) {};
\node at (2,1) (j2) {};
\node at (-2,-1) (j3) {};
\node at (2,-1) (j4) {};
\draw [->,line width=2pt] (-3,-0.1) .. controls (2,-7) and (2,7) .. (-3,0.1);
\draw[decorate,
decoration={snake}] (j1) -- (j2);
\draw[decorate,
decoration={snake}] (j3) -- (j4);
\node at (-3.5,0) (i1) {};
\node at (3.5,0) (i2) {};
\node at (0,-3.5) (i3) {};
\node at (0,3.5) (i4) {};
%\draw [->] (i1) -- (i2);
%\draw [->,] (i3) -- (i4);
\end{tikzpicture}
    \end{center}
  \end{minipage}
  \hfill
  \begin{minipage}[t]{.47\textwidth}
    \begin{center}\begin{tikzpicture}[scale=0.8]
\clip (-3,-3.5) rectangle (5,3.5);
\draw (-2.6,-3) rectangle (4.8,3.4);
\node at (-1.4,3) (k1) {$\boxed{\varepsilon=1}$};
\node at (3.5,2.9) (k5) {$\boxed{q^{-2iu_1}}$};
\node at (3.8,1.2) (m1) {$|x_1^-|=1$};
\node at (3.8,-1.2) (m1) {$|x_1^+|=1$};
\node at (0.1,0.1) (j1) {};
\node at (3,1) (j2) {};
\node at (0.1,-0.1) (j3) {};
\node at (3,-1) (j4) {};
\draw [->,line width=2pt] (-2,-0.1) .. controls (3,-7) and (3,7) .. (-2,0.1);
\draw[decorate,
decoration={snake}] (j1) -- (j2);
\draw[decorate,
decoration={snake}] (j3) -- (j4);
\node at (-3.5,0) (i1) {};
\node at (3.5,0) (i2) {};
\node at (0,-3.5) (i3) {};
\node at (0,3.5) (i4) {};
%\draw [->] (i1) -- (i2);
%\draw [->,] (i3) -- (i4);
\end{tikzpicture}
    \end{center}
  \end{minipage}
   \begin{minipage}[t]{.47\textwidth}
    \begin{center}
\begin{tikzpicture}[scale=0.8]
\clip (-4,-3.5) rectangle (3.8,3.5);
\draw (-3.8,-3) rectangle (3.6,3.4);
\node at (-2,3) (k1) {$\boxed{\varepsilon=-1}$};
\node at (2.8,3) (k5) {$\boxed{u_1}$};
\node at (2.5,1.5) (m1) {$|x_1^-|=1$};
\node at (2.5,-1.5) (m1) {$|x_1^+|=1$};
\node at (-3,-2) (k2) {${\cal R}_{-2,0}$};
\node at (-2.5,-1) (n2) {};
\draw [->] (k2) -- (n2);
\node at (-3,2) (k3) {${\cal R}_{0,0}$};
\node at (-2.5,1) (n3) {};
\draw [->] (k3) -- (n3);
\node at (2,0.4) (k4) {${\cal R}_{-1,0}$};
\node at (0.7,0) (n4) {};
\draw [->] (k4) -- (n4);
\node at (2,1) (j1) {};
\node at (-2,1) (j2) {};
\node at (2,-1) (j3) {};
\node at (-2,-1) (j4) {};
\draw [->,line width=2pt] (-3,0.1) .. controls (2,7) and (2,-7) .. (-3,-0.1);
\draw[decorate,
decoration={snake}] (j1) -- (j2);
\draw[decorate,
decoration={snake}] (j3) -- (j4);
\node at (-3.5,0) (i1) {};
\node at (3.5,0) (i2) {};
\node at (0,-3.5) (i3) {};
\node at (0,3.5) (i4) {};
%\draw [->] (i1) -- (i2);
%\draw [->,] (i3) -- (i4);
\end{tikzpicture}
    \end{center}
  \end{minipage}
  \hfill
  \begin{minipage}[t]{.47\textwidth}
    \begin{center}\begin{tikzpicture}[scale=0.8]
\clip (-3,-3.5) rectangle (5,3.5);
\draw (-2.6,-3) rectangle (4.8,3.4);
\node at (3.5,2.9) (k5) {$\boxed{q^{-2iu_1}}$};
\node at (-1.4,3) (k1) {$\boxed{\varepsilon=-1}$};
\node at (3.8,1.2) (m1) {$|x_1^-|=1$};
\node at (3.8,-1.2) (m1) {$|x_1^+|=1$};
\node at (0.1,0.1) (j1) {};
\node at (3,1) (j2) {};
\node at (0.1,-0.1) (j3) {};
\node at (3,-1) (j4) {};
\draw [->,line width=2pt] (-2,0.1) .. controls (3,7) and (3,-7) .. (-2,-0.1);
\draw[decorate,
decoration={snake}] (j1) -- (j2);
\draw[decorate,
decoration={snake}] (j3) -- (j4);
\node at (-3.5,0) (i1) {};
\node at (3.5,0) (i2) {};
\node at (0,-3.5) (i3) {};
\node at (0,3.5) (i4) {};
%\draw [->] (i1) -- (i2);
%\draw [->,] (i3) -- (i4);
\end{tikzpicture}
    \end{center}
  \end{minipage}
  \caption{\small The contours for analytic continuation $z_1\to z_1+\varepsilon\omega_2$.
The top/bottom figures show $\varepsilon=1$/$-1$. The left figures show the situation in the $u_1$ plane in the string limit while the right figures show the situation near the relativistic limit in the $q^{-2iu_1}$ plane where the left-hand branch points nearly touch at the origin. In the magnon theory $\theta_1=-\pi u_1/k$ and the top/bottom figures correspond to $\IM\theta_1$ decreasing/increasing so that in the relativistic limit the analytic continuation corresponds to $\theta_1\to\theta_1-i\varepsilon\pi$. The opposite is true in the mirror theory. The left figures show the sheets ${\cal R}_{n,0}$ that the contour crosses when starting on ${\cal R}_{0,0}$.}
\label{f6}
\end{figure}

The particular analytic continuation needed in the magnon theory was made clear in \cite{Arutyunov:2009kf,Volin:2009uv,Vieira:2010kb}. It corresponds to
$\varepsilon=1$ in our notation and involves the anti-clockwise contour in the 
$u_1$ plane as shown in the top left figure \ref{f6}. We also show how this contour generalizes in the  $q^{-iu_1}$-plane when $q\neq1$. Note that with this choice the analytic continuation involves the sheets ${\cal R}_{0,0}\to{\cal R}_{1,0}\to{\cal R}_{2,0}$. 
The two choices for the analytic continuation $\varepsilon=\pm1$  in the  string limit are illustrated in the left-hand figures of \ref{f6} which shows the contours for the analytic continuation in the $u_1$-plane.

The two choices $\varepsilon=\pm1$ lead to S-matrices with different analytic properties and so the choice matters. In particular, the difference is 
not the usual kind of CCD ambiguity that always appears in the S-matrices of integrable theories. The string world-sheet theory apparently requires $\varepsilon=+1$ and we now argue that once the 
S-matrices are $q$-deformed and become interpolating S-matrices then only the mirror theory is consistent with the relativistic limit. Since this is a key observation, let us lay out the logic of the argument seriatim:

(i) First of all, we have fixed the way that $u$, $x^\pm$ or $z$ relates to the relativistic rapidity in the $g\to\infty$ limit by taking into account the position of the $s$-channel bound-state pole of the S-matrix which occurs at 
\EQ{\label{pref}
\text{magnon:}&\qquad x_1^+=x_2^-\ ,\qquad u_1-u_2=-i\ ,\\
\text{mirror:}&\qquad x_1^-=x_2^+,\qquad~\, u_1-u_2=i\ ,
}
on the sheet ${\cal R}_{0,0}$. Assuming that the bound state poles continue to lie on the physical sheet in the interpolating theory, 
 \eqref{pp2} is fixed by the condition that both correspond to $\theta\equiv\theta_1-\theta_2=\frac{\pi i}k$. 

(ii) It is clear from figure \ref{f6} that the analytic continuation $z_1\to z_1+\varepsilon\omega_2$ with $\varepsilon=1$ involves crossing the cut $|x_1^+|=1$ first and then $|x_1^-|=1$ and therefore since $x_1^\pm=x(u_1\pm\frac i2)$ the imaginary part of $u_1$ is increasing. Indeed, in top right-hand diagram of  figure \ref{f6} the motion is counter-clockwise and since $q^{-2iu_1}=e^{2\pi u_1/k}$ it must be that $\IM u_1$ is increasing. Consequently, in the relativistic limit with the choice $\varepsilon=1$, $\IM\theta_1$ is decreasing, $\theta_1\to\theta_1-i\pi$, in the magnon theory and increasing , $\theta_1\to\theta_1+i\pi$, in the mirror theory. Note that in the relativistic limit the left-hand branch points nearly touch at the origin forcing the contour to encircle the origin. This implies that moving round the contour is equivalent to wrapping the non-trivial cycle of the $u$/rapidity-cylinder. The opposite is true with the choice $\varepsilon=-1$.
To summarize, the relation is 
\begin{equation}\boxed{\begin{split}
\text{magnon:}&\qquad z\to z+\varepsilon\omega_2 \qquad\Longleftrightarrow\qquad\theta\to\theta-i\varepsilon\pi\ ,\\
\text{mirror:}&\qquad z\to z+\varepsilon\omega_2 \qquad\Longleftrightarrow\qquad\theta\to\theta+i\varepsilon\pi\ .\end{split}}
\label{bp1}
\end{equation}
In the relativistic limit, the crossing equation \eqref{css2} becomes
\EQ{
\sigma(\theta\mp i\varepsilon\pi)\sigma(\theta)=\frac{\cosh(\frac\theta2)\sinh(\frac\theta2\pm\frac{i\pi}{2k})}
{\sinh(\frac\theta2)\cosh(\frac\theta2\mp\frac{i\pi}{2k})}\ ,
\label{nrw}
}
where the upper/lower sign is for the magnon and mirror cases, respectively. This is the correct relativistic crossing equation only if the left-hand side is $\sigma(\theta)\sigma(\theta+i\pi)$; that is $\varepsilon=-1$, in the magnon case, and  $\varepsilon=1$, in the mirror case.

If the conventional dressing phase $\sigma$ is the solution of \eqref{css2} with $\varepsilon=1$, let us denote the solution with $\varepsilon=-1$ as $\widehat\sigma $. If we take \eqref{css2} with $\varepsilon=-1$ and then shift $z_1\to z_1+\omega_2$  we have
\EQ{
\widehat\sigma (z_1,z_2)\widehat\sigma (z_1+\omega_2,z_2)=\frac1{U_2^{2}p(z_2,z_1+\omega_2)}
=\frac{x_2^-+\xi}{x_2^++\xi}\cdot \frac{\frac1{x_1^-}-x_2^+}{\frac1{x_1^-}-x_2^-}\cdot\frac{1-\frac{x_1^+}{x_2^+}}{1-\frac{x_1^+}{x_2^-}}\ .
}
Then multiplying this with \eqref{css2} with $\varepsilon=1$ and using \eqref{trick} gives
\EQ{
\sigma(z_1+\omega_2,z_2)&\sigma(z_1,z_2)
\widehat\sigma (z_1+\omega_2,z_2)\widehat\sigma (z_1,z_2)\\ &
=q^2\cdot\frac{x_2^+}{x_2^-}\cdot\left(\frac{x_2^-+\xi}{x_2^++\xi}\right)^2\cdot\frac{1-\frac{x_1^-}{x_2^+}}{1-\frac1{x_1^+x_2^-}}\cdot
\frac{1-\frac1{x_1^-x_2^+}}{1-\frac{x_1^+}{x_2^-}}
\ .
}
There are two solutions of this equation for $\widehat\sigma $ in terms of $\sigma$ which are consistent with unitarity. We choose the solution which will imply that if $\sigma$ and its inverse are analytic on ${\cal R}_{0,0}$ then so is $\widehat\sigma $, and vice-versa; namely,
\EQ{
\widehat\sigma (z_1,z_2)=
\frac{U_1V_2}{U_2V_1}\cdot\frac{1-\frac1{x_1^-x_2^+}}{1-\frac1{x_1^+x_2^-}}\cdot\frac1{\sigma(z_1,z_2)}\ .
\label{rpm}
}
Notice that the pre-factor here has no poles or zeros on ${\cal R}_{0,0}$.
We can then define two additional S-matrices by
choosing  $\widehat\sigma $ as the dressing factor rather than $\sigma$. 

Out of the four possible S-matrices built from the  magnon or mirror bootstrap with either
$\sigma$ or $\widehat\sigma$ as the dressing phase, only the mirror S-matrix with $\sigma$ and the magnon S-matrix with $\widehat\sigma $ satisfy the correct crossing equation of relativistic S-matrix theory in the limit $g\to\infty$. This is summarized in table \ref{GTable}.

\begin{table}[ht]
\begin{center}
\begin{tabular}{ccccc}
\toprule
 type & dressing & $s$-channel &   bound states & relativistic crossing \\
\otoprule
magnon & $\sigma$ & $x_1^+=x_2^-$ & $\langle n,0\rangle$ & No\\
mirror & $\sigma$ & $x_1^-=x_2^+$ & $\langle 0,n\rangle$ & Yes\\
magnon & $\widehat\sigma $ & $x_1^+=x_2^-$ & $\langle n,0\rangle$ & Yes\\
mirror & $\widehat\sigma $ & $x_1^-=x_2^+$ & $\langle 0,n\rangle$ & No\\
\bottomrule
\end{tabular}
\end{center}
\caption{\small A summary of the data for the four possible S-matrices. The top two lines are the conventional magnon and mirror theories, whereas the last two lines are the same S-matrices but with the alternative dressing factor $\widehat\sigma $. Note that only the middle two lines are consistent with relativistic crossing symmetry in the $g\to\infty$ limit and so that includes the mirror theory but apparently not the ordinary magnon theory.}
\label{GTable}
\end{table}

We can illustrate the difference between the four S-matrices by considering the $\msu(2)$ sector in the string limit. The magnon/mirror S-matrix takes the form \eqref{xpp}
while the new S-matrix involves replacing $\sigma$ with $\widehat\sigma$ which can then be written in terms of $\sigma$ using \eqref{rpm}, to give
\EQ{
\widehat S_{\msu(2)}(z_1,z_2)=\sigma(z_1,z_2)^2\cdot\frac{x_1^--x_2^+}{x_1^+-x_2^-}\cdot
\frac{1-\frac1{x_1^+x_2^-}}{1-\frac1{x_1^-x_2^+}}\ .
\label{ot2}
}
The magnon version of this manifests the bound-state
$s$-channel pole at $x_1^+=x_2^-$, or $u_1-u_2=-i$ and also the  $t$-channel pole $x_1^-=1/x_2^+$, or $u_1-u_2=i$, on the sheet ${\cal R}_{-2,0}$. For the mirror version, both these poles lie off the physical sheet and the $\msu(2)$ element has no bound-state poles.

\section{Constructing the Dressing Phase}

We now turn to the solution for the dressing phase $\sigma $ in the $q$ deformed theory. The story of the dressing phase in the magnon theory is long and interesting and summarized nicely in the review article \cite{Vieira:2010kb}. A conjecture for an integral form of the phase \cite{Dorey:2007xn} was shown in \cite{Arutyunov:2009kf} to explicitly satisfy the crossing equation \eqref{css2} (for $\varepsilon=1$). A
constructive derivation of the factor then appeared in \cite{Volin:2009uv}.  We will draw heavily on these later two references as well as on the review \cite{Vieira:2010kb}.

To start with we will assume that in the $q$ deformed theory $\sigma(z_1,z_2)$ has the same factorization as in the magnon S-matrix \cite{Arutyunov:2006iu}:\footnote{Here, we are using the conventions of \cite{Arutyunov:2009kf} (but with $g\to2g$). In \cite{Volin:2009uv,Vieira:2010kb} $\chi(x_1,x_2)$ is defined with the opposite sign.}
\EQ{
\sigma (z_1,z_2)=\exp i\big[\chi(x_1^+,x_2^+)-\chi(x_1^-,x_2^+)-\chi(x_1^+,x_2^-)+\chi(x_1^-,x_2^-)\big]\ ,
\label{dcp}
}
with $\chi(x_1,x_2)=-\chi(x_2,x_1)$ in order to ensure that the unitarity constraint \eqref{unit1} is satisfied. 

The first step of \cite{Volin:2009uv,Vieira:2010kb} is to show that the quantity $\chi(u_1,u_2)\equiv\chi(x_1=x(u_1),x_2=x(u_2))$ satisfies a Riemann-Hilbert problem.
As a function of the $u_i$, $\chi(u_1,u_2)$ inherits the branch cuts of $x_i=x(u_i)$ corresponding to $u_i\in{\EuScript C}$. The Riemann-Hilbert problem takes the 
form
\EQ{
&\chi(u_1+\epsilon,u_2+\epsilon)+\chi(u_1+\epsilon,u_2-\epsilon)\\ &+\chi(u_1-\epsilon,u_2+\epsilon)+\chi(u_1-\epsilon,u_2-\epsilon)=i
\log\Theta(u_1,u_2)\ ,
\label{mg3}
}
where $u_i\in{\EuScript C}$ and $\epsilon$ is an infinitesimal such that $u_i\pm \epsilon$ lie on either side of the cut. 
The solution of such a problem is then straightforward. Firstly, if we have the simpler problem
\EQ{
F(u+\epsilon)+F(u-\epsilon)=G(u)\ ,
\label{sp2}
}
then the solution can be written as\footnote{In order to show that it satisfies \eqref{sp2}, one finds, using $x(u+\epsilon)=1/x(u-\epsilon)$ for $u\in\EuScript C$,  that the two terms on the left-hand side of \eqref{sp2} almost cancel but leave an integral around the pole at $w=u$ with a residue that is precisely $G(u)$.} 
\EQ{
F(u)=K_u\ast G(u)\ ,
}
where we have defined
\EQ{
K_{u}\ast G(u)=\frac{2\pi}k\int_{{\EuScript C}+\epsilon}\frac{dw}{2\pi i}\cdot\frac{x(u)-\frac1{x(u)}}{x(w)-\frac1{x(w)}}\cdot\frac1{1-q^{2i(w-u)}}G(w)\ .
\label{bxy}
}
Using this integral kernel we can write the solution of the Riemann-Hilbert problem \eqref{mg3} as
\EQ{
\chi(u_1,u_2)=i\, K_{u_1}\ast K_{u_2}\ast\log\Theta(u_1,u_2)\ .
\label{bxx}
}
Then we use the identity
\EQ{
\xi\left(\frac{g}{\tilde g}\right)^2\cdot\frac{x(u)-\frac1{x(u)}}{q^{-2iw}-q^{-2iu}}
=-1+\frac{x(w)}{x(w)-x(u)}+\frac{\frac1{x(w)}}{\frac1{x(w)}-x(u)}\ .
}
The first term here is independent of $u$ and does not contribute in the combination \eqref{dcp}. We then change variables from $w$ to $z=x(w)$, in the second term, and to $z=1/x(w)$ in the third term, to end up with the 
integral form \cite{Dorey:2007xn,Volin:2009uv,Vieira:2010kb}
\EQ{
\chi(x_1,x_2)=i\oint_{|z|=1}\frac{dz}{2\pi i}\,\frac1{z-x_1}\oint_{|z'|=1}\frac{dz'}{2\pi i}\,\frac1{z'-x_2}
\log\Theta(u(z),u(z'))\ ,
\label{fdd2}
}
up to terms which do not contribute to the dressing phase due to the combination of functions in \eqref{dcp}. The solution of the problem therefore boils down to specifying the 
kernel $\Theta(u_1,u_2)$ in the Riemann-Hilbert problem. Note that the unitarity constraint \eqref{unit1} requires
\EQ{
\Theta(u_1,u_2)\Theta(u_2,u_1)=1\ .
}
In the string limit, the kernel has to satisfy the equation \cite{Volin:2009uv,Vieira:2010kb}
\EQ{
\Theta_\text{mag}(u_1,u_2)^{-D+D^{-1}}=\left(\frac{x(u_1)-\frac1{x(u_2)}}{\sqrt{x(u_1)}}
\cdot\frac{x(u_1)-x(u_2)}{\sqrt{x(u_1)}}\right)^{D+D^{-1}}\ ,
\label{xpl}
}
where $D=\exp(\frac i2\partial_{u_1})$ is an operator that acts on functions to shift $u_1$: $Df(u_1)=f(u_1+\frac i2)$. This yields the difference equation
\EQ{
\frac{\Theta_\text{mag}(u_1-\frac i2,u_2)}{\Theta_\text{mag}(u_1+\frac i2,u_2)}=\Big(u_1-u_2-\frac i2\Big)\Big(u_1-u_2+\frac i2\Big)\ .
\label{mg1}
}
The solution of these conditions is not unique, but it was argued that constraints on the analytic properties of the dressing phase lead to the particular solution \cite{Volin:2009uv,Vieira:2010kb}
\EQ{
\Theta_\text{mag}(u_1,u_2)=\frac{\Gamma(1+iu_1-iu_2)}{\Gamma(1-iu_1+iu_2)}\ .
\label{mg2}
}
 
For the interpolating theory the equation for the kernel is still \eqref{xpl} but now the map $x(u)$ is modified and we can use this to motivate a solution for $q\neq1$. Using the modified map, the difference equation \eqref{mg1} now becomes 
\EQ{
\frac{\Theta(u_1-\frac i2,u_2)}{\Theta(u_1+\frac i2,u_2)}=\left(q^{-2iu_1+1}-q^{-2iu_2}\right)\left(q^{-2iu_1-1}-q^{-2iu_2}\right)\ ,
}
up to multiplicative factors which do not affect the result. 
This suggests that the solution should be similar to \eqref{mg2}
with gamma functions replaced by $q$-gamma functions.
\EQ{
\Theta(u_1,u_2)=
\frac{\Gamma_{q^2}(1+iu_1-iu_2)}{\Gamma_{q^2}(1-iu_1+iu_2)}\ ,
\label{it2}
}
where the conventional $q$-gamma function satisfies the basic identity
\EQ{
\Gamma_{q^2}(1+x)=\frac{1-q^{2x}}{1-q^2}\Gamma_{q^2}(x)\ .
\label{apa}
}
The expression \eqref{it2} solves the difference equation, again up to multiplicative factors
that we have not kept track of. Our ultimate justification for taking \eqref{it2} comes with the direct proof of crossing later in this section.

The conventional gamma function has the integral representation
\EQ{
\log\,\Gamma(1+x)=\int_0^\infty\frac{dt}t\,\frac{e^{-tx}-x(e^{-t}-1)-1}{e^t-1}\ ,
\label{ja1}
}
valid for $\RE\,x>-1$. We can write a similar kind of representation for the $q$-gamma function with $q=e^{i\pi/k}$:\footnote{We do not know whether this representation has appeared in the literature before.}
\EQ{
&\log\Gamma_{q^2}(1+x)=\frac{i\pi x(x-1)}{2k}\\ &\qquad+\int_0^\infty\frac{dt}t\,\frac{e^{-tx}-e^{(x-k+1)t}
-x(e^{-t}-1)(1+e^{(2-k)t})+e^{(1-k)t}-1}{(e^{t}-1)(1-e^{-kt})}\ .
\label{cll}
}
This representation is valid in the strip $-1< \RE x<k$ (with $k>1$) and returns to \eqref{ja1} in the large $k$ limit. We can then use \eqref{apa} to analytically continue to other regions of the complex $x$-plane. Note that the terms which are constant, linear and quadratic in $x$ in \eqref{cll} do not contribute to the dressing phase. This is either because they cancel when taking the logarithm of \eqref{it2}, or because the integral in \eqref{fdd2}, for a kernel which is only a function of $u_1$ (or $u_2$), vanishes as the contour of the $z'$ (or $z$) integral can be shrunk to nothing.

To summarize, our solution for the dressing phase in the interpolating theory is
\EQ{
\chi(x_1,x_2)=i\oint_{|z|=1}\frac{dz}{2\pi i}\,\frac1{z-x_1}\oint_{|z'|=1}\frac{dz'}{2\pi i}\,\frac1{z'-x_2}
\log\frac{ \Gamma_{q^2}(1+iu(z)-iu(z'))}
{ \Gamma_{q^2}(1-iu(z)+iu(z'))}\ .
\label{fdd3}
}

We can find a useful representation of $\chi(x_1,x_2)$ through its large $x_i$ expansion:
\EQ{
\chi(x_1,x_2)=\sum_{r,s=1}^\infty\frac{c_{rs}(g,k)}{x_1^rx_2^s}\ ,
}
where
\EQ{
c_{rs}(g,k)=i\int_0^{2\pi}\frac{d\phi}{2\pi}\int_0^{2\pi}\frac{d\phi'}{2\pi}\,e^{ir\phi+is\phi'}
\log\frac{ \Gamma_{q^2}(1+iu(e^{i\phi})-iu(e^{i\phi'}))}
{ \Gamma_{q^2}(1-iu(e^{i\phi})+iu(e^{i\phi'}))}\ .
}
Notice that the image of $|z|=1$ in the $u$ plane lies along the real axis and so we can use the integral representation for the $q$-gamma function \eqref{cll} to write
\EQ{
c_{rs}(g,k)=i\int_0^\infty\frac{dt}t\,\frac{1+e^{(1-k)t}}{(e^t-1)(1-e^{-k t})}
\big(L_r(t)L_s(-t)-L_r(-t)L_s(t)\big)\ ,
}
where we have defined
\EQ{
L_r(t)=\int_0^{2\pi}\frac{d\phi}{2\pi}\,e^{ir\phi}\big(1+\xi^2+2\xi\cos\phi\big)^{-ikt/2\pi}\ .
}
In the limit $k\to\infty$, $\xi\to 2\pi g/k$, and so
\EQ{
L_r(t)\longrightarrow\int_0^{2\pi}\frac{d\phi}{2\pi}\,e^{ir\phi}e^{-2igt\cos\phi}
=e^{-i\pi r/2} J_r(2gt)
}
and we recover the well-known series expansion of the magnon dressing phase with coefficients \cite{Beisert:2006ez}
\EQ{
c_{rs}(g)_\text{mag}=2\sin\frac{\pi(r-s)}2\int_0^\infty\frac{dt}t\,\frac{J_r(2gt)
J_s(2gt)}{e^t-1}\ .
}

\vspace{0.5cm}\noindent{\bf Direct proof of crossing}

In this section we consider the analytic structure of the dressing phase and show explicitly that it satisfies the crossing equation. Our approach mirrors very closely that of   
\cite{Arutyunov:2009kf} and we draw on all of the techniques developed there generalized appropriately for the interpolating theory. In view of this we find it very convenient to use the same notation. 

In order that the crossing equation \eqref{css2} is satisfied, we start on the sheet ${\cal R}_{0,0}$ for which we take 
\EQ{
{\cal R}_{0,0}\,:\qquad \chi(x_1,x_2)=\Phi(x_1,x_2)\ ,
}
where the right-hand side is the integral expression in \eqref{fdd3}.
In the region ${\cal R}_{0,0}$, $|x_1^\pm|>1$ and $|x_2^\pm|>1$, the integral representation manifests that the dressing phase $\sigma(z_1,z_2)$ and its inverse are  analytic in this region.

\begin{figure}[t]
  \hfill
  \begin{minipage}[t]{.32\textwidth}
    \begin{center}
\includegraphics[width=0.99\textwidth]{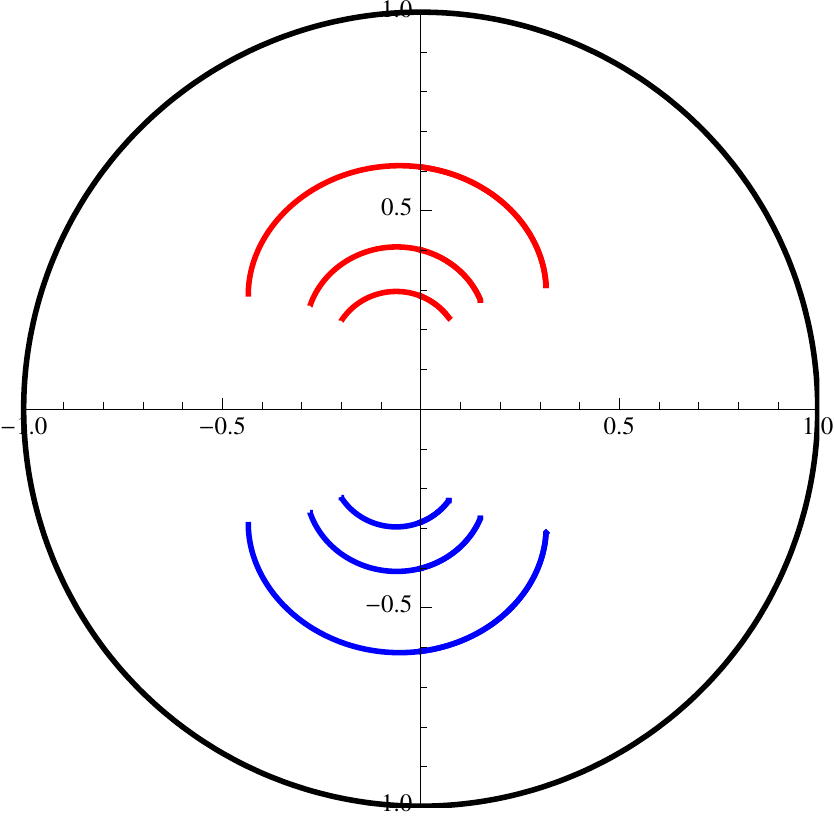}
    \end{center}
  \end{minipage}
  \hfill
  \begin{minipage}[t]{.32\textwidth}
    \begin{center}
\includegraphics[width=0.99\textwidth]{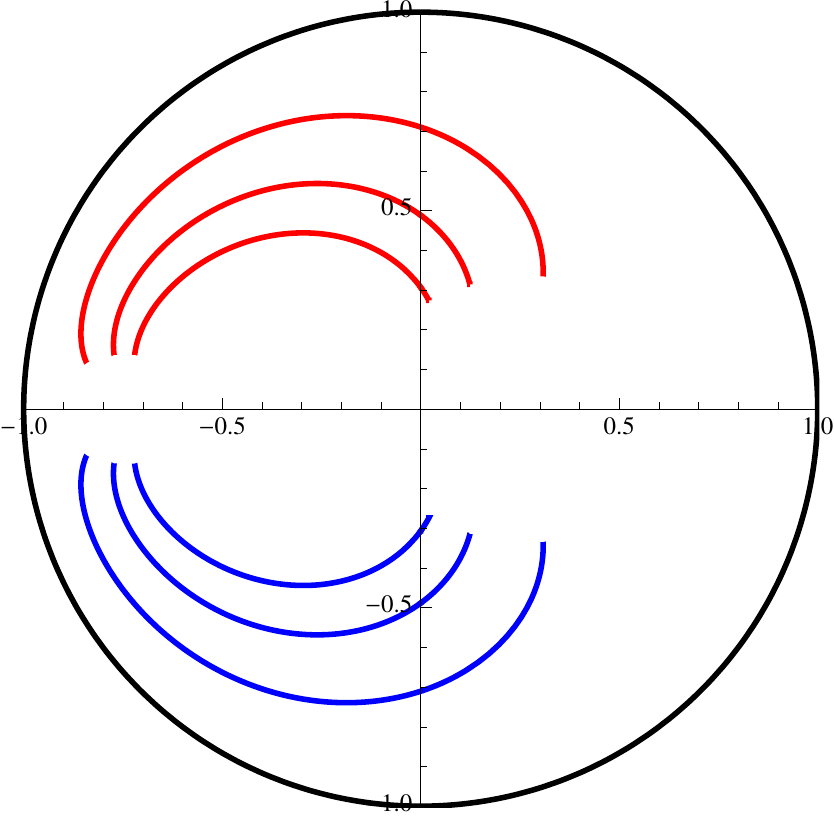}
    \end{center}
  \end{minipage}
  \hfill
    \begin{minipage}[t]{.32\textwidth}
    \begin{center}
\includegraphics[width=0.99\textwidth]{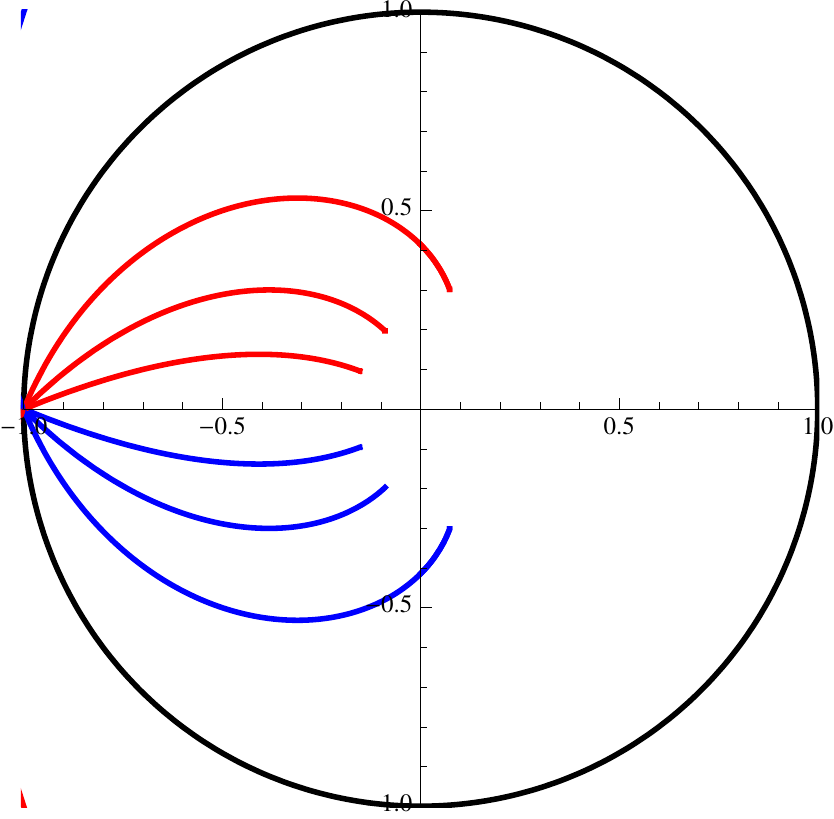}
    \end{center}
  \end{minipage}
  \hfill
\caption{The cuts $\text x^{(n)}_\pm$ in the $x$-plane for $n=1,2,3$ ($+$ in blue, $-$ in red) with $n$ increasing as the cuts move closer to the origin. The unit circle is shown in black. The three plots show the cuts in the (left) string limit with $g=1$, $k=50$; (middle) intermediate regime $k=20$, $g=3$; (right) relativistic limit $k=8$, $g=15$. In the relativistic limit $g\to\infty$ the left-hand branch points coalesce at $x=-1$ and the region of interest is the circular neighbourhood of $x=-1$.} 
\label{f5}
\end{figure}

Let us now consider the crossing equation \eqref{css2} with $\varepsilon =1$. This involves analytically continuing ${\cal R}_{0,0}\to{\cal R}_{1,0}\to{\cal R}_{2,0}$, as illustrated in  figure \ref{f6}. The first change of sheet here involves crossing the curve $|x_1^+|=1$ which is the boundary of the validity of the integral representation. Crossing this curve involves an additive modification of the integral representation for the functions $\chi(x_1^+,x_2^\pm)$:
\EQ{
{\cal R}_{1,0}\,:\qquad \chi(x_1^+,x_2^\pm)&=\Phi(x_1^+,x_2^\pm)-\Psi(x_1^+,x_2^\pm)\ ,\\
\chi(x_1^-,x_2^\pm)&=\Phi(x_1^-,x_2^\pm)\ .
}
Note that $\chi(x_1^-,x_2^\pm)$ is not modified since we still have $|x_1^-|>1$ on the sheet ${\cal R}_{1,0}$.
$\Psi(x_1,x_2)$ is easily calculated as in the magnon case by picking up the residue of  the $z$ integral around the pole at $z=x_1$ to give
\EQ{
\Psi(x_1,x_2)=i\oint_{|z|=1}\frac{dz}{2\pi i}\frac1{z-x_2}\log\frac{\Gamma_{q^2}(1+iu(x_1)-iu(z))}{
\Gamma_{q^2}(1-iu(x_1)+iu(z))}\ .
\label{gwe}
}
The function $\Psi(x_1,x_2)$ is itself a branched function of $x_1$ with cuts that are solutions of $u(x_1)=u(z)\pm in$, for $n\in\mathbb Z\neq0$ and $u(z)\in{\EuScript C}$; that is
\EQ{
x_1+\frac1{x_1}+\xi+\frac1\xi=q^{\pm2n}\left(z+\frac1z+\xi+\frac1\xi\right)\ ,\qquad z\in\{e^{i\phi},\ 0\leq\phi<2\pi\}
} 
with $|x_1|<1$. If we choose to define $x(u)$ on the sheet with $|x(u)|>1$ then 
the cuts are defined as 
\EQ{
\text x^{(n)}_\pm=\Big\{x=\frac1{x(u\pm in)}\ ,\ u\in{\EuScript C}\Big\}\ .
}
The cuts are illustrated in figure \ref{f5} for three choices of $(g,k)$.

The fact that there are multiple cuts on the sheet ${\cal R}_{1,0}$ means that one must be careful to specify the path for the analytic continuation. As discussed at length in \cite{Arutyunov:2009kf}, the path must be chosen so that $x_1^+$ crosses 
the cut $\text x^{(1)}_+$ first. This means that as we 
perform the second part of the analytic continuation ${\cal R}_{1,0}\to{\cal R}_{2,0}$,
which involves moving from a sheet with $|x_1^-|>1$ to $|x_1^-|<1$, the variable $x_1^+$ also crosses the cut
$\text x^{(1)}_+$ since $x_1^+\in \text x^{(1)}_+$ implies $|x_1^-|=1$.
So as we pass across $|x_1^-|=1$ from ${\cal R}_{1,0}\to{\cal R}_{2,0}$ two things happen: firstly, since $|x_1^-|=1$ lies at the boundary of the region of validity of the integral representation of $\Phi(x_1^-,x_2^\pm)$, $\chi(x_1^-,x_2^\pm)$ picks up an additional contribution of $-\Psi(x_1^-,x_2^\pm)$ and secondly 
$\Psi(x_1^+,x_2^\pm)$ picks up its own discontinuity as in \eqref{se1}, with $z(x_1)=x_1^-$, as $x_1^+$ crosses the cut $\text x^{(1)}_+$.  Putting this together, we have
\EQ{
{\cal R}_{2,0}\,:\qquad \chi(x_1^+,x_2^\pm)&=\Phi(x_1^+,x_2^\pm)-\Psi(x_1^+,x_2^\pm)-i\log\frac{x_2^\pm-\frac1{x_1^-}}{x_2^\pm-x_1^-}\ ,\\
\chi(x_1^-,x_2^\pm)&=\Phi(x_1^-,x_2^\pm)-\Psi(x_1^-,x_2^\pm)\ .
\label{vb1}
}

\begin{figure}
  \hfill
  \begin{minipage}[t]{.32\textwidth}
    \begin{center}
\begin{tikzpicture}[scale=0.67]
\draw (-3,-3) rectangle (4,3);
\node at (-2,2.3) (i1) {${\cal R}_{0,0}$};
\node at (2.5,2.2) (m1) {$|x_1^-|=1$};
\node at (2.5,-2.2) (m2) {$|x_1^+|=1$};
\draw [->] (m1) -- (2,1.1);
\draw [->] (m2) -- (2,-1.1);
\draw[decorate,
decoration={snake}] (0.28,0.11) -- (2.77,1.15);
\draw[decorate,
decoration={snake},color=red] (0.28,-0.11) -- (2.77,-1.15);
\draw [-,line width=2pt] (-2,-0.3) .. controls (-1,-2) and (1,-2) .. (1.84,-0.71);
\end{tikzpicture}
    \end{center}
  \end{minipage}
  \hfill
  \begin{minipage}[t]{.32\textwidth}
    \begin{center}
\begin{tikzpicture}[scale=0.67]
\draw (-3,-3) rectangle (4,3);
\node at (-2,2.3) (i1) {${\cal R}_{1,0}$};
\draw[decorate,
decoration={snake},color=blue] (0.28,0.11) -- (2.77,1.15);
\draw[decorate,
decoration={snake},color=red] (0.28,-0.11) -- (2.77,-1.15);
\draw[decorate,
decoration={snake}] (0.11,0.28) -- (1.15,2.77);
\draw[decorate,
decoration={snake}] (0.11,-0.28) -- (1.15,-2.77);
\draw[decorate,
decoration={snake}] (-0.11,0.28) -- (-1.15,2.77);
\draw[decorate,
decoration={snake}] (-0.11,-0.28) -- (-1.15,-2.77);
\filldraw[black] (-1.41,1.41) circle (2pt);
\filldraw[black] (-1.41,-1.41) circle (2pt);
\filldraw[black] (-1.84,0.76) circle (2pt);
\filldraw[black] (-1.84,-0.76) circle (2pt);
\draw [-,line width=2pt] (1.84,-0.71) .. controls (2,-0.3) and (2,0.3) .. (1.84,0.84);
\end{tikzpicture}
    \end{center}
  \end{minipage}
  \hfill
  \begin{minipage}[t]{.32\textwidth}
    \begin{center}
\begin{tikzpicture}[scale=0.67]
\draw (-3,-3) rectangle (4,3);
\node at (-2,2.3) (i1) {${\cal R}_{2,0}$};
\node at (-3,0) {};
\draw[decorate,
decoration={snake},color=blue] (0.28,0.11) -- (2.77,1.15);
\draw[decorate,
decoration={snake}] (0.28,-0.11) -- (2.77,-1.15);
\draw [<-,line width=2pt] (-2,0.3) .. controls (-1,2) and (1,2) .. (1.84,0.84);
\end{tikzpicture}
    \end{center}
  \end{minipage}
  \hfill
\caption{The cut structure of $\sigma(z_1,z_2)$ in the $q^{-2iu}$ plane on the sheets ${\cal R}_{0,0}$, ${\cal R}_{1,0}$ and ${\cal R}_{2,0}$. The red and blue cuts are identified. Note that the black cuts on the sheet ${\cal R}_{1,0}$ corresponding to the cuts $\text{x}^{(n)}_+$, $n=2,3,\dots$ going anti-clockwise from the cut $|x_1^-|=1$ (which is $\text{x}^{(1)}_+$), and $\text{x}^{(n)}_-$, $n=1,2,\ldots$ going clockwise from the cut $|x_1^+|=1$, of $\chi(x_1^+,x_2^\pm)$ become out of reach in the relativistic limit when all the inner branch points  coalesce at the origin and the outer ones go to infinity. Also shown is the path for the analytic continuation $z_1\to z_1+\omega_2$.} 
\label{f10}
\end{figure}
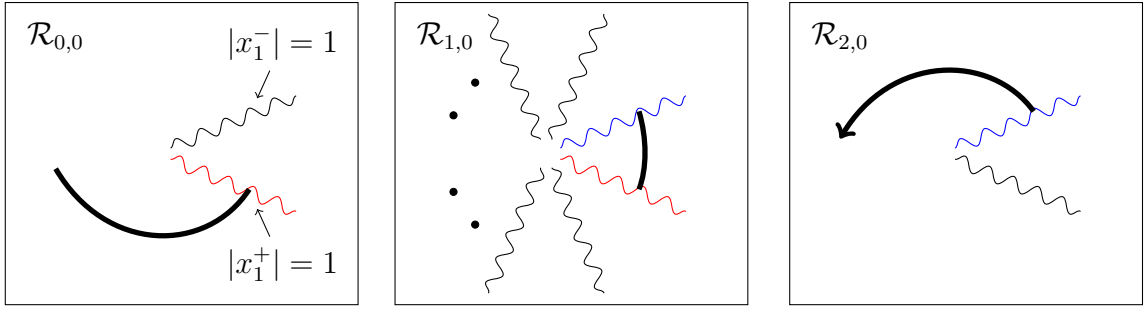

We are now in a position to prove that the crossing relation \eqref{css2} is satisfied for $(z_1, z_2)\in{\cal R}_{0,0}$ with the choice of analytic continuation described above. Using the identity
\EQ{
\Phi(x_1,x_2)+\Phi\Big(\frac1{x_1},x_2\Big)=\Phi(0,x_2)\ ,
}
proved in Appendix \ref{a2},
and \eqref{vb1}, we find that for $(z_1,z_2)\in{\cal R}_{0,0}$ 
\EQ{
&-i\log\left[\sigma (z_1+\omega_2,z_2)\sigma (z_1,z_2)\right]=
\Psi\Big(\frac1{x_1^-},x_2^+\Big)-\Psi\Big(\frac1{x_1^+},x_2^+\Big)\\ &\qquad\qquad\qquad\qquad+\Psi\Big(\frac1{x_1^+},x_2^-\Big)-\Psi\Big(\frac1{x_1^-},x_2^-\Big)
 -i\log\frac{x_2^+-x_1^-}{x_2^+-\frac1{x_1^-}}\cdot\frac{x_2^--\frac1{x_1^-}}{x_2^--x_1^-}\ .
\label{jj2}
}
Then using the identity
\EQ{
&\Psi\Big(\frac1{x_1^-},x_2^+\Big)-\Psi\Big(\frac1{x_1^+},x_2^+\Big)+\Psi\Big(\frac1{x_1^+},x_2^-\Big)-\Psi\Big(\frac1{x_1^-},x_2^-\Big)\\
&\qquad\qquad\qquad\qquad=-i\log\frac{1-\frac1{x_1^-x_2^+}}{1-\frac1{x_1^-x_2^-}}\cdot\frac{1-\frac1{x_1^+x_2^+}}{1-\frac1{x_1^+x_2^-}}+i\log\frac{1+\frac\xi{x_2^+}}{1+\frac\xi{x_2^-}}\ ,
}
which follows from \eqref{id2} proved in Appendix \ref{a2}, \eqref{jj2} becomes precisely the crossing equation \eqref{css2}. The cut structure of $\sigma(z_1,z_2)$ is illustrated in figure \ref{f10}. 

We have shown above that the crossing equations are satisfied between regions ${\cal R}_{0,0}$ and ${\cal R}_{2,0}$. As explained in \cite{Arutyunov:2009kf}, in order to complete the proof of crossing one has to show that it is satisfied between regions ${\cal R}_{0,1}$ and ${\cal R}_{2,1}$. We leave this straightforward generalization to the reader.

\vspace{0.5cm}
\noindent{\bf The Relativistic Soliton Limit}

In the relativistic limit, the branch points of ${\EuScript C}$, $u_\pm$, move to $\mp\infty$. The cut ${\EuScript C}$ then corresponds to the infinite line $\IM u=0$ (modulo $k$).  The situation in the $x$ plane is shown in figure \ref{f5}: the left branch points of the cuts $\text x^{(n)}_\pm$ all coalesce on $x=-1$ and the limit is then determined by the behaviour in the neighbourhood of $x=-1$. This has the important consequence that in the relativistic limit the 
dressing phase becomes a meromorphic function, as is required by relativistic S-matrix theory. In particular, the branch cuts at $x_1^+\in \text x^{(n)}_+$ for $n>1$ on the sheet ${\cal R}_{1,0}$ are now hidden behind the cut $\text x^{(1)}_+$ 
and the analytic continuation described in the last section becomes unique. The behaviour of the cuts near the relativistic limit is shown in figure \ref{f10}. In particular, this makes clear why the additional cuts on the sheet ${\cal R}_{1,0}$ become out of reach in the relativistic limit. Note that the analytic continuation $z_1\to z_1+\omega_2$ corresponding to
${\cal R}_{0,0}\to{\cal R}_{1,0}\to{\cal R}_{2,0}$ corresponds to $\theta_1\to\theta_1-i\pi$ in the relativistic limit for the magnon theory, and $\theta_1\to\theta_1+i\pi$, for the mirror.

In order to take the relativistic limit of the dressing phase directly, it is useful to go back to the expression \eqref{bxx} with \eqref{bxy}. The map $x(u)$ becomes \eqref{frq} in the relativistic limit. Given that the kernel of \eqref{fdd3} can be expressed as an integral using \eqref{cll}, then up to terms which do not contribute to the dressing phase
\EQ{
\log\Theta(u_1,u_2)=\int_0^\infty\frac{dt}t\cdot\frac{(e^{-i(u_1-u_2)t}-e^{i(u_1-u_2)t})(1+e^{(1-k)t})
}{(e^{t}-1)(1-e^{-kt})}+\cdots\ ,
\label{cll2}
}
and we need to evaluate the integral
\EQ{
\II(u,t)&=\frac{2\pi}k
\int_{{\EuScript C}}\frac{dw}{2\pi i}\cdot\frac{x(u)-\frac1{x(u)}}{x(w)-\frac1{x(w)}}\cdot\frac{e^{-iwt}}{1-q^{2i(w-u)}}\\ &\underset{g\to\infty}\longrightarrow
\frac{2\pi}k\int_{-\infty}^\infty\frac{dw}{2\pi i}\cdot\frac{e^{-iwt}}{e^{\pi(w-u)/k}-e^{-\pi(w-u)/k}}\ .
}
Note that in the relativistic limit the contour $\EuScript C$ becomes the real axis.
We can evaluate the integral by completing the contour at infinity and picking up the residues taking into account that $|x(u)|>1$ implies $0<\IM u<k$. The poles are at $w=u+ikn$, with $n\in\mathbb Z$. Completing the contour at infinity in either the upper- or lower-half planes gives 
\EQ{
\II(u,t)=\frac{e^{-iut}}{1+e^{kt}}\ .
}
Hence, with $u_i=\mp k\theta_i/\pi$, for the magnon and mirror theories, respectively,
\EQ{
\chi(\theta_1,\theta_2)&= i\int_0^\infty\frac{dt}t\cdot\frac{1+e^{(1-k)t}}{(e^t-1)(1-e^{-kt})}\Big(\II(u_1,t)\II(u_2,-t)-\II(u_1,-t)\II(u_2,t)\Big)\\
&=\mp\frac 12\int_0^\infty\frac{dt}t\cdot\frac{\cosh((k-1)t)\sin(\frac{2k\theta t}\pi)}{\sinh(2kt)\sinh(t)\cosh(kt)}\ ,
}
where to get the last line we have scaled $t\to2t$ and set $\theta=\theta_1-\theta_2$. The latter means that relativistic invariance is recovered in the limit. Note that this integral is divergent at $t=0$, however, the singular term cancels in the combination \eqref{dcp}, that is
\EQ{
\sigma (\theta)=\exp i\big[2\chi(\theta)-\chi(\theta-\tfrac{i\pi}k)-\chi(\theta+\tfrac{i\pi}k)\big]\ ,
\label{dcp2}
} 
with $\chi(\theta)\equiv\chi(\theta_1,\theta_2)$. It is easy to show that the latter expression
yields 
\EQ{
\sigma (\theta)=\exp\left[\pm2i\int_0^\infty \frac{dt}t\,\frac{\sinh(t)\cosh((k-1)t)\sin(\frac{2k\theta t}\pi)}{\cosh(kt)\sinh(2kt)}\right]\ ,
\label{fww2}
}
with the upper/lower sign for the magnon and mirror cases, respectively. Notice that the magnon and mirror cases are related by $\theta\to-\theta$:
\EQ{
\sigma(\theta)\Big|_\text{mirror}=\sigma(-\theta)\Big|_\text{magnon}=\sigma(\theta)^{-1}\Big|_\text{magnon}\ .
\label{bcz}
}

It is reasonably simple to write the dressing factor \eqref{fww2} in terms of an infinite product of gamma functions; choosing the magnon case,
\EQ{
\sigma(\theta)\Big|_\text{magnon}=\frac{\cosh(\tfrac{\theta}{2}+\tfrac{i\pi}{2k})}{\cosh(\tfrac{\theta}{2}-\tfrac{i\pi}{2k})}\cdot\frac{\rho(-\theta)\rho(\theta-i\pi)}{\rho(\theta)\rho(-\theta-i\pi)}\ ,
\label{gqq}
} 
where
\EQ{
\rho(\theta)=\prod_{\ell=0}^\infty\frac{\Gamma(-\frac\theta{2i\pi}+\frac12+\ell)
\Gamma(-\frac\theta{2i\pi}+\frac1{2k}+1+\ell)}
{\Gamma(-\frac\theta{2i\pi}+1+\ell)
\Gamma(-\frac\theta{2i\pi}-\frac1{2k}+\frac12+\ell)} \ .
}
For the mirror case we simply use \eqref{bcz}.
These expressions manifest the fact the $\sigma(\theta)$ has no poles or zeros in the region $|\IM\theta|<\pi$ in either the magnon or mirror cases.

Taking the relativistic limit of \eqref{rpm} and using \eqref{gqq}, the alternative dressing factor
in the relativistic limit is 
\EQ{
\widehat\sigma (\theta)\Big|_\text{magnon}=\frac{\rho(\theta)\rho(-\theta-i\pi)}{\rho(-\theta)\rho(\theta-i\pi)}\ ,
\label{gqq2}
}
which means that we can write
\EQ{
\widehat\sigma (\theta)=\exp\left[\pm2i\int_0^\infty \frac{dt}t\,\frac{\sinh(t)\cosh((k+1)t)\sin(\frac{2k\theta t}\pi)}{\cosh(kt)\sinh(2kt)}\right]\ ,
\label{fww}
}
which is valid in the region $|\IM\theta|<\pi-\frac\pi k$, with  the upper/lower sign for the magnon and mirror cases, respectively.
Contrary to $\sigma(\theta)$, the alternative dressing factor $\widehat\sigma (\theta)$
has a zero at $\theta=i\pi-\frac{i\pi}k$, which gives the $t$-channel pole of the magnon type S-matrix, and a pole at $\theta=-i\pi+\frac{i\pi}k$. These are then swapped over in the mirror case according to $\theta\to-\theta$.

The magnon type S-matrix with $\widehat\sigma $ is precisely the S-matrix constructed in \cite{Hoare:2011nd}. The exact relation to the quantity ${\cal F}(\theta)$ defined in that reference is
\EQ{
\frac1{\widehat\sigma (\theta)}=\frac{\cosh(\tfrac\theta{2})\sinh(\tfrac{\theta}{2}+\tfrac{i\pi}{2k})}
{\sinh(\tfrac{i\pi}{2k})}\cdot
\frac{\cosh(\tfrac{\theta}{2}-\tfrac{i\pi}{2k})}{\cosh(\tfrac{\theta}{2}+\tfrac{i\pi}{2k})}\cdot{\cal F}(\theta)\ .
\label{rsf}
}

It is interesting that the S-matrices of the magnon and mirror type are simply related by the parity transformation $\theta\to-\theta$.\footnote{The language ``magnon" and ``mirror" here refer to the behaviour of the S-matrix in the string limit and not the relativistic limit. In particular, the relativistic limits of the magnon and mirror theories do not appear to be identical as one would expect in a relativistic theory, although this is only a puzzle if the ``mirror map" is well defined for the interpolating S-matrix . We address this issue at the end of the next section.
}
 Out of the four possible S-matrices it is only the two shown in table \ref{GTable} that are consistent with relativistic crossing symmetry. This includes, of course the S-matrix constructed in \cite{Hoare:2011nd}. The transformation $\theta\to-\theta$ has the effect of changing the bound states from the symmetric representations $\langle n,0\rangle$ to the anti-symmetric representations $\langle0,n\rangle$, as we discuss more fully in the next section.

\section{Bound-State Processes}\label{s6}

The S-matrices that we have constructed have a complicated analytic structure: there are poles, zeros and branch points. In the relativistic limit things are simpler: the branch points disappear and the S-matrix is a meromorphic function. In this limit a special role is played by the {\it physical strip\/}, the region $0\leq \IM\theta\leq\pi$, which corresponds to the {\it physical sheet\/}  in the usual $s$-parameterization familiar from S-matrix theory in $3+1$-dimensions. Simple poles in this region correspond to bound states propagating in either the $s$- or $t$-channel. Higher poles correspond to anomalous thresholds that become branch points in higher dimensions.\footnote{Sometimes anomalous thresholds can give rise to simple poles. A classic example of this occurs in the S-matrix of sine-Gordon theory  clearly explained in \cite{Dorey:1996gd}.}
Unfortunately, we do not know the analogue of the physical strip---or sheet---in non-relativistic S-matrix theory.\footnote{Note that in the magnon theory, the region ${\cal R}_{0,0}$ is sometimes called the ``physical region" however this is not the ``physical sheet" of the non-relativistic theory; for instance the $t$-channel pole is on the sheet ${\cal R}_{2,0}$ and this must be on the ``physical sheet".}

\FIG{
\begin{tikzpicture} [scale=1,line width=1.5pt,inner sep=2mm,
place/.style={circle,draw=blue!50,fill=blue!20,thick},proj/.style={circle,draw=red!50,fill=red!20,thick}]
\node at (1.5,1.5) [proj] (p1) {};
\node at (0,0) (i1) {$|i,z_1\rangle$};
\node at (3,0) (i2) {$|j,z_2\rangle$};
\draw[-]  (i2) -- (p1);
\draw[-]  (i1) -- (p1);
\node at (1.5,3) [proj] (p2) {};
\node at (0,4.5) (i3) {$|k,z_2\rangle$};
\node at (3,4.5) (i4) {$|l,z_1\rangle$};
\draw[<-]  (i3) -- (p2);
\draw[<-]  (i4) -- (p2);
\draw[-]  (p1) -- (p2);
\node at (-1,2.2) {$\boxed{{\cal R}_{0,0}:\quad x_1^+=x_2^-}$};
\node at (2.9,2.2) {$\boxed{\theta=\frac{\pi i}k}$};
\begin{scope}[xshift=5.5cm,yshift=1.5cm]
\node at (1.5,1.5) [proj] (p3) {};
\node at (0,0) (j1) {$|i,z_1\rangle$};
\node at (4.5,0) (j2) {$|j,z_2\rangle$};
\node at (0,3) (j3) {$|k,z_2\rangle$};
\node at (4.5,3) (j4) {$|l,z_1\rangle$};
\draw[<-]  (j3) -- (p3);
\draw[-]  (j1) -- (p3);
\node at (3,1.5) [proj] (p4) {};
\draw[-]  (j2) -- (p4);
\draw[<-]  (j4) -- (p4);
\draw[-]  (p3) -- (p4);
\node at (2.2,3.3) {$\boxed{\theta=-i\varepsilon\pi-\frac{\pi i}k}$};
\node at (2.2,-1) {$\boxed{{\cal R}_{2\varepsilon,0}:\quad x_1^-=\frac1{x_2^+}}$};
\end{scope}
\end{tikzpicture}
\caption{\small The bound-state processes giving rise to simple poles of the magnon S-matrix on the sheet as indicated. On the left is the $s$-channel process and on the right the $t$-channel process. Note that the $t$-channel pole occurs on the sheet ${\cal R}_{2\varepsilon,0}$. Also shown are the rapidity differences in the relativistic limit. For the mirror case the $s$-channel pole is at $x_1^-=x_2^+$ and the $t$-channel pole $x_1^+=1/x_2^-$.}
\label{f11}
}

In the string limit, we know the energy and momentum of states and so there are conventional arguments that identify the physical poles. Note that the mirror theory has different energy and momentum from the magnon theory and so the physical poles will be different.
For instance, if we take the magnon S-matrix, the pole at $x_1^+=x_2^-$ on the sheet ${\cal R}_{0,0}$ is physical and corresponds to the bound state $\langle1,0\rangle$ in the $s$-channel. As we $q$-deform, the interpolating S-matrix has this pole as is apparent in \eqref{pcm}: each $R$-matrix has a pole and $Z(z_1,z_2)$ has a zero at $x_1^+=x_2^-$ so overall there is a simple pole as expected. Note that the dressing phases $\sigma$ or $\widehat\sigma $ have no poles or zeros on the sheet ${\cal R}_{0,0}$. 
Crossing symmetry then implies that there should be a $t$-channel pole at $x_1^-=1/x_2^+$ on the sheet ${\cal R}_{2\varepsilon,0}$ corresponding to the bound state  $\langle1,0\rangle$. 
The question is whether the S-matrix has a simple pole at this point. The answer must be yes since the crossing equation \eqref{csm} relates the S-matrix on the sheet ${\cal R}_{0,0}$ on the left-hand side to the S-matrix on the sheet ${\cal R}_{2\varepsilon,0}$ on the right-hand side. Then the fact that the left-hand side has a simple pole at $x_1^+=x_2^-$ implies that the right-hand side must have a simple pole at $x_1^+=x_2^-$. This then implies that
the  S-matrix must have a  simple pole at $x_1^-=1/x_2^+$ on the sheet ${\cal R}_{2\varepsilon,0}$ corresponding to a bound-state in the $t$-channel.
As we have argued, it is only with the dressing factor $\widehat\sigma $ that we get a consistent S-matrix in the relativistic limit. These poles are illustrated in figure \ref{f11}.
For the mirror theory, the physical sheet changes and now the 
$s$-channel pole is at $x_1^-=x_2^+$ and the $t$-channel pole at $x_1^+=1/x_2^-$.
The $s$- and $t$-channel poles corresponding to the bound states are summarized in table \ref{t2}.

\begin{table}[ht]
\begin{center}
\begin{tabular}{ccccccc}
\toprule
 type & dressing &   bound states & \multicolumn{2}{c}{$s$-channel} &   \multicolumn{2}{c}{$t$-channel}\\
&&& generic & rel. limit &generic &rel. limit\\
\otoprule
magnon & $\sigma$ &$\langle n,0\rangle$ & $x_1^+=x_2^-$& $\theta=\frac{i\pi}k$& $x_1^-=\frac1{x_2^+}$ &$\theta=-i\pi-\frac{i\pi}k$ \\ 
mirror & $\sigma$ &$\langle 0,n\rangle$ & $x_1^-=x_2^+$& $\theta=\frac{i\pi}k$& $x_1^+=\frac1{x_2^-}$ &$\theta=i\pi-\frac{i\pi}k$ \\ 
magnon & $\widehat\sigma $ &$\langle n,0\rangle$ & $x_1^+=x_2^-$& $\theta=\frac{i\pi}k$& $x_1^-=\frac1{x_2^+}$ &$\theta=i\pi-\frac{i\pi}k$ \\ 
mirror & $\widehat\sigma $ &$\langle 0,n\rangle$ & $x_1^-=x_2^+$& $\theta=\frac{i\pi}k$& $x_1^+=\frac1{x_2^-}$ &$\theta=-i\pi-\frac{i\pi}k$ \\ 
\bottomrule
\end{tabular}
\end{center}
\caption{\small This summarizes the positions of the $s$-and $t$-channel poles for the four possible S-matrices including the relativistic limit. Note that the $t$-channel pole is in the wrong place for the magnon S-matrix based on $\sigma$ and the mirror S-matrix based on $\widehat\sigma $ since relativistic crossing symmetry is violated for these theories.}
\label{t2}
\end{table}

Other poles of the interpolating S-matrix should correspond to anomalous thresholds and some work in this direction appears in \cite{Dorey:2007xn}. In particular, note that these ``DHM poles" are located on sheets that are reached by crossing the cuts $
\text{x}^{(n)}_+$ in figure \ref{f10} and these sheets become out of reach in the relativistic limit. Therefore the DHM poles are not associated to anomalous thresholds in the relativistic theory.\footnote{Note in this regard that the DHM pole on the sheet reached by crossing the particular cut $\text{x}^{(1)}_+$ is actually the $t$-channel pole and {\it not\/} an anomalous threshold.}

We know from either the magnon or mirror limits that the theories contain bound states transforming in (a product of 2 copies) of the atypical representations 
\EQ{
\text{magnon:}\qquad \langle a-1,0\rangle\ ,\qquad
\text{mirror:}\qquad \langle 0,a-1\rangle\ ,
}
$a=1,2,\ldots$,
of the deformed algebra $U_q(\mh)$. For these representations, the shortening condition \eqref{sht} generalizes to:
\EQ{
[C]_q^2-PK=\Big[\frac a2\Big]_q^2\ .
}
As for the fundamental representation $a=1$, one can introduce parameters $x^\pm$, but now with a modified dispersion relation\footnote{This generalized dispersion relation was also written down independently in \cite{deLeeuw:2011jr}.}
\EQ{
q^{-a}\Big(x^++\frac1{x^+}\Big)-q^a\Big(x^-+\frac1{x^-}\Big)=(q^a-q^{-a})\Big(\xi+\frac1\xi\Big)\ .
\label{p11m}
}
The shortening condition is equivalent to the dispersion relation if the central charges are given as in \eqref{cch} where $U$ and $V$ are now given more generally by
\EQ{
U^2=q^{-a}\frac{x^++\xi}{x^-+\xi}=q^a\frac{\frac1{x^-}+\xi}{\frac1{x^+}+\xi}\ ,\qquad
V^2=q^{-a}\frac{\xi x^++1}{\xi x^-+1}=q^a\frac{\frac\xi{x^-}+1}{\frac\xi{x^+}+1}\ .
}
In terms of the map $x(u)$, we have
\EQ{
x^\pm=x\Big(u\pm \frac{ia}2\Big)\ .
}
In principle the S-matrix elements of the bound states can be found by using the bootstrap equations, as employed in the relativistic limit in \cite{Hoare:2011nd}. However, the R-matrix on which the S-matrix is based can also be deduced on purely algebraic grounds as explained in \cite{deLeeuw:2011jr}. The possible 3-point vertices are illustrated in figures \ref{f1}. These couplings are relevant only in the mirror theory where there is a good relativistic limit in which case $\VV{a}$  are the modules for the anti-symmetric representations $\langle 0,a-1\rangle$.\footnote{For the magnon theory with $\widehat\sigma $ one has to replace $x^\pm\leftrightarrow x^\mp$ and $\VV{a}$ are the symmetric representations.} 
These vertices mean that the general S-matrix element $S_{ab}(z_1,z_2)$ should have four bound-state poles, two in the $s$- and two in the $t$-channel corresponding to bound-states $\VV{a+b}$ and  $\VV{|a-b|}$ in each case. 
\FIG{
\begin{tikzpicture} [line width=1.5pt,inner sep=2mm,
place/.style={circle,draw=blue!50,fill=blue!20,thick},proj/.style={circle,draw=red!50,fill=red!20,thick}]
\node at (2,2) [proj] (p1) {};
\node at (0.4,0.4) (i1) {$\VV{a}(x_1^\pm)$};
\node at (3.6,0.4) (i2)  {$\VV{b}(x_2^\pm)$};
\node at (2,4) (i3) {$\VV{a+b}(x_3^-=x_2^-,x_3^+=x_1^+)$};
\draw[-]  (i1) -- (p1);
\draw[-]  (i2) -- (p1);
\draw[-]  (i3) -- (p1);
\node at (2,0.3) {$\boxed{x_1^-=x_2^+}$};
\node at (8,2) [proj] (p1) {};
\node at (5,2) (hel) {$\xrightarrow{~~~g\to\infty~~~}$};
\node at (6.4,0.4) (i1) {$\VV{a}(\theta+\frac{i\pi b}{2k})$};
\node at (9.6,0.4) (i2)  {$\VV{b}(\theta-\frac{i\pi a}{2k})$};
\node at (8,4) (i3) {$\VV{a+b}(\theta)$};
\draw[-]  (i1) -- (p1);
\draw[-]  (i2) -- (p1);
\draw[-]  (i3) -- (p1);
\end{tikzpicture}
\vspace{0.9cm}
\begin{tikzpicture} [line width=1.5pt,inner sep=2mm,
place/.style={circle,draw=blue!50,fill=blue!20,thick},proj/.style={circle,draw=red!50,fill=red!20,thick}]
\node at (2,2) [proj] (p1) {};
\node at (0.2,0.4) (i1) {$\VV{a}(x_1^\pm)$};
\node at (3.8,0.4) (i2)  {$\VV{b}(x_2^\pm)$};
\node at (2,4) (i3) {$\VV{a-b}(x_3^-=x_1^-,x_3^+=\frac1{x_2^-})$};
\draw[-]  (i1) -- (p1);
\draw[-]  (i2) -- (p1);
\draw[-]  (i3) -- (p1);
\node at (2,0.3) {$\boxed{x_1^+=\frac1{x_2^+}}$};
\node at (8,2) [proj] (p1) {};
\node at (5,2) (hel) {$\xrightarrow{~~~g\to\infty~~~}$};
\node at (6.4,0.4) (i1) {$\VV{a}(\theta+\frac{i\pi b}{2k})$};
\node at (9.6,0.4) (i2)  {$\VV{b}(\theta-i\pi+\frac{i\pi a}{2k})$};
\node at (8,4) (i3) {$\VV{a-b}(\theta)$};
\draw[-]  (i1) -- (p1);
\draw[-]  (i2) -- (p1);
\draw[-]  (i3) -- (p1);
\end{tikzpicture}
\caption{\small The three point vertices for incoming states in $\VV{a}(x_1^\pm)\otimes \VV{b}(x_2^\pm)$ showing the rapidities in the relativistic limit $g\to\infty$ for the mirror theory. Those in the second line assume that $a>b$. If $b>a$ then the incoming states have $x_1^+=\frac1{x_2^+}$ and the outgoing state space  is $\VV{b-a}(x_3^+=\frac1{x_1^-},x_3^-=x_2^-)$ and in the relativistic limit the incoming state space are $\VV{a}(\theta+i\pi-\frac{i\pi b}{2k})\otimes \VV{b}(\theta-\frac{i\pi a}{2k})$ with an outgoing state space is $\VV{b-a}(\theta)$.}
\label{f1}
}

The theory with $k\in\mathbb Z$, so that $q$ is a root of unity $q^{2k}=1$, is particularly interesting because the spectrum of bound states is naturally truncated $\langle a-1,0\rangle$ or $\langle0,a-1\rangle$, $a=1,2,\ldots,k$. In particular, this provides a natural regularization of the infinite spectrum of states in the magnon or mirror theory which are obtained in the limit $k\to\infty$.

\vspace{0.5cm}
\noindent{\bf A Possible Resolution of the Crossing Puzzle}

The picture that we have established is slightly unsatisfactory in the sense that the original magnon S-matrix does not seem to have a good relativistic limit when $q$-deformed. 
One possible way to rectify this feature arises from our lack of knowledge of the physical sheet of the S-matrix. It could be that at some intermediate values of $g$ and $k$, as we move from a neighbourhood of the string limit to one of the relativistic limits, the $s$-channel pole at $x_1^+=x_2^-$ moves off the physical sheet and the pole at $x_1^-=x_2^+$ moves on to the physical sheet.
This would mean that at that point in parameter space, the bound states $\langle a-1,0\rangle$, for $a>1$, become unbound and $\langle0,a-1\rangle$ become bound. If we {\em assume} this is the case, then to take the relativistic limit of the q deformed magnon S-matrix with dressing factor $\s$, the logic in part (i) on page \pageref{pref} implies $\o = +\tfrac{\pi}{k}u$. Consequently the crossing relation becomes the correct relativistic one.  
Then in the relativistic limit the magnon and mirror S-matrices become identical, as expected in a relativistic theory. This would be consistent with the idea that the double Wick rotation has an action on the interpolating S-matrix and not just on the string and relativistic limits. Further work will be needed to clarify if this possibility actually is correct.

\section{Discussion}

In this paper we have constructed S-matrices that correspond to a $q$ deformation of the S-matrix of the magnons of the string world-sheet in $\text{AdS}_5\times S^5$. The main work involved finding the dressing factor that ensures the unitarity and crossing symmetry of the theory. We pointed out an ambiguity in defining the crossing equation in the non-relativistic S-matrix theory that gives rise to two different dressing factors $\sigma$ and $\widehat\sigma $. This allows for the existence of four distinct S-matrices: the magnon and mirror S-matrices with either $\sigma$ or $\widehat\sigma $ as dressing factor. Once these S-matrices are embedded in the larger theory incorporating the $q$ deformation, then consistency with the relativistic limit fixes the ambiguity. The deformation of the original magnon S-matrix does not satisfy crossing symmetry in the relativistic limit but the mirror S-matrix does. This means that one can define a TBA system for the interpolating mirror theory and this would be interesting to investigate.

\section*{Acknowledgements}

\noindent
BH is supported by EPSRC and would like to thank Dmytro Volin for many valuable discussions and Arkady Tseytlin for enjoyable collaborations and useful discussions on related topics.

\noindent
TJH is supported in part by the STFC grant ST/G000506/1 and
would like to thank the TH Division, CERN for hospitality
while some of this work was carried out.

\noindent
JLM is supported in part by MICINN (FPA2008-01838 and 
FPA2008-01177), Xunta de Galicia (Consejer\'\i a de Educaci\'on and INCITE09.296.035PR), the Spanish Consolider-Ingenio 2010
Programme CPAN (CSD2007-00042), and FEDER.

\vspace{1cm}

\appendix
\appendixpage

\section{Uniformizing the Rapidity Torus}\label{a1}

In order to find the rapidity torus, we note that
the algebraic equation \eqref{p11} defines an elliptic curve. This can be shown directly by writing the algebraic relation in Weierstrass form, generalizing \cite{Janik:2006dc} for the magnon case. A more direct approach which also leads to the rapidity torus is to find an 
explicit parameterization of $x^\pm$ in terms of Jacobi elliptic functions. This was done partially in \cite{Beisert:2008tw}. The curve turns out to have a modulus 
\EQ{
\kappa=4i\frac{g^2}{\tilde g}=4ig\sqrt{1-g^2(q-q^{-1})^2}=4ig\sqrt{1+4g^2\sin^2\frac\pi k}\ ,
}
which has the well-known string limit $\kappa=4ig$. We also define the conventional squared modulus $m=\kappa^2=-16g^4/\tilde g^2$.\footnote{We use the Mathematica convention that the elliptic function $K=K(m)$ and $K'=K(1-m)$.} After some trial and error (and more knowledge of elliptic functions than is healthy) one can find the expressions for $x^\pm$ in terms of Jacobi elliptic functions for a torus of modulus $\kappa$,\footnote{Our expressions are consistent with the expression (2.67) of \cite{Beisert:2008tw} for the original variables $\tilde x^\pm$.}
\EQ{
x^+(z)=\frac{\tilde g}{2g^2}\cdot\frac{q^2 i\tau \dn(z)-i(q^2-2g^2(q-1)^2(q+1))-8ig^4\tilde g^{-2}q(q-1)\sn^2(z)}
{2q^2i\tau \cn(z)\sn(z)+q(q-1)+2(2g^2(q-1)^2(q+1)-q^2)\sn^2(z)}\ ,
}
where $\tau=\sqrt{1-4g^2(q^{1/2}-q^{-1/2})^2}$, with $x^-(z)$ given by the same expression with $q\to q^{-1}$ and $i\to-i$. In the string limit $q\to1$, we have
\EQ{
x^\pm(z)=\frac{\dn(z)-1}{4g\sn(z)}\big(\cn(z)\mp i\sn(z)\big)\ ,
}
which are the known expressions in that case.

A convenient choice of periods of the rapidity torus is $2\omega_1=4K(m)$ and $2\omega_2=4iK(1-m)-4K(m)$ since, assuming $g$ is real and positive and $q$ is a complex phase, then $\omega_1$ is real and $\omega_2$ is purely imaginary. The crossing symmetry antipode operation corresponds to a shift by half a period positively or negatively in the $\omega_2$ direction:
\EQ{
x^\pm(z\pm\omega_2)=\frac1{x^\pm(z)}\ .
}

In the relativistic limit $g\to\infty$,  we can use the $s\to0$ asymptotic forms
\EQ{
K(-1/s)\to\frac{\sqrt s}2\log(16/s)\ ,\qquad
K(1+1/s)\to\frac{\sqrt s}2\big(\pi-i\log(16/s)\big)\ ,
}
to show that at leading order
\EQ{
\omega_1\to \sqrt s\log(16/s)\ ,\qquad\omega_2\to \pi i\sqrt s\ ,
}
with $s=(64g^4\sin^2\tfrac{\pi}{k})^{-1}$.
In this limit, both periods vanish, however $\omega_2$ vanishes faster. This 
suggests a re-scaling $z=-\sqrt s\theta$, which in the relativistic limit has a periodicity of 
$\theta\sim \theta+2\pi i$ in the $\w_2$ direction
and divergent periodicity in the $\w_1$ direction.  
In this limit,
\EQ{
\sn(z)\to -\sqrt{s}\sinh(\theta)\ ,\qquad \dn(z)\to\cosh(\theta)\ ,
}
along with $\cn(z)\to1$, and so one finds the relation \eqref{frq} along with \eqref{pp2}.

\section{The Magnon S-matrix}\label{app2}

In this appendix, we survey some of the literature regarding the magnon S-matrix and crossing symmetry in order to establish our conventions and show they relate to other works. This is not intended to be a comprehensive review of the literature.
In particular, we will concentrate on the S-matrix in the $\msu(2)$ sector that is for  $|\phi^a\phi^a;z_1\rangle\otimes\ket{\phi^a\phi^a;z_2}\to|\phi^a\phi^a;z_2\rangle\otimes\ket{\phi^a\phi^a;z_1}$.

We begin with Janik's paper which established how to formulate crossing symmetry \cite{Janik:2006dc}. In this work, the S-matrix elements are based on the $R$-matrix of Beisert \cite{hep-th/0511082} which are precisely the $q\to1$ limit of the $R$-matrix of 
 \cite{Beisert:2008tw} written in \eqref{jjs}, \eqref{eqnsb} (but without the $U_iV_i$ factors).
The S-matrix in the $\msu(2)$ sector is taken as $S_0A_{12}^2$ that is
\EQ{
S_{\msu(2)}(z_1,z_2)=S_0(z_1,z_2)\left(\frac{x_2^+-x_1^-}{x_2^--x_1^+}\right)^2\ .
\label{qw1}
}
Janik's crossing equation can then be written as\footnote{Note we have  $\omega_1\leftrightarrow\omega_2$ relative to Janik.}
\EQ{
S_0(z_1+\omega_2,z_2)S_0(z_1,z_2)=\left(\frac{x_1^+-x_2^+}{x_1^--x_2^+}\cdot\frac{1-\frac1{x_1^+x_2^-}}{1-\frac1{x_1^-x_2^-}}\right)^2\ .
\label{pw2}
}

Now we turn to Beisert's paper \cite{hep-th/0606214}. It uses the same conventions as Janik above and writes the S-matrix element in the $\msu(2)$ sector as
\EQ{
S_{\msu(2)}(z_1,z_2)=\frac1{\sigma(z_1,z_2)^2}\cdot\frac{x_1^--x_2^+}{x_1^+-x_2^-}\cdot\frac{1-\frac1{x_1^-x_2^+}}{1-\frac1{x_1^+x_2^-}}\ .
\label{qw2}
}
 The dressing factor then satisfies the crossing equation---eq.~9---(in our notation)
\EQ{
\sigma(z_1+\omega_2,z_2)\sigma(z_1,z_2)=\frac{x_2^-}{x_2^+}\cdot
\frac{x_1^--x_2^+}{x_1^--x_2^-}\cdot\frac{1-\frac1{x_1^+x_2^+}}{1-\frac1{x_1^+x_2^-}}\ ,
\label{pw1}
}
which is precisely \eqref{pw2} re-written in terms of $\sigma$. 
The form of the crossing equation \eqref{pw2} appears as eq.~3.59 and 3.60 in \cite{Beisert:2006qh} although without the square since there is only one $R$-matrix factor in that reference.

The crossing equation \eqref{pw1} appears in the review article \cite{Arutyunov:2009ga} for the S-matrix in the $\msu(2)$ sector---eq.~3.94---written as (in our notation)\footnote{In this reference $S_0=S_{\msu(2)}$.}
\EQ{
S_{\msu(2)}(z_1,z_2)&=\frac1{\sigma(z_1,z_2)^2}\cdot\frac{x_1^+x_2^-}{x_1^-x_2^+}\cdot\frac{x_1^--x_2^+}{x_1^+-x_2^-}\cdot\frac{1-\frac1{x_1^-x_2^+}}{1-\frac1{x_1^+x_2^-}}\\
&=\frac1{\sigma(z_1,z_2)^2}\cdot\frac{x_1^+x_2^-}{x_1^-x_2^+}\cdot\frac{u_1-u_2-i}{u_1-u_2+i}\ .
}
We have  not written the factors involving exponentials of $p_i$ as these can be removed by a simple re-definition. Note the factors of $x_i^\pm$ which can be re-defined away for the single element $\msu(2)$ as above in \eqref{qw1} and \eqref{qw2}, but then this would affect other elements of the S-matrix. We prefer to leave these factors in the definition of $S_{\msu(2)}$ since they appear naturally in the $q\to1$ limit of the $R$-matrix in \eqref{jjs}. The crossing equation 
eq.~3.118 written in terms of $\sigma$ is then identical to \eqref{pw1}.

In many references the $\msu(2)$ S-matrix written is the inverse of the one written here. Using unitarity \eqref{run} this is equivalent to a parity transformation $z_1\leftrightarrow z_2$. For example, in \cite{Arutyunov:2006iu} the $\msu(2)$ S-matrix written in eq.~3 with $\mathfrak s=1$ is the inverse of \eqref{qw1} and \eqref{qw2} with $\sigma^2\to\sigma$. However, the crossing equation eq.~12 with $x_j=x_1$ and $x_k=x_2$ is identical to \eqref{pw1}.

\section{Some Useful Identities}\label{a2}

In this appendix, we establish some identities that are needed in the proof of crossing symmetry. For the most part, these identities are either identical or simple generalizations of those established in \cite{Arutyunov:2009kf}.

The first identity takes the identical form to the magnon case:
\EQ{
\Phi(x_1,x_2)+\Phi\Big(\frac1{x_1},x_2\Big)=\Phi(0,x_2)\ ,
\label{id1}
}
where $\Phi(x_1,x_2)$ is the integral defined in \eqref{fdd3}. This is easily proved by changing variable $z\to z^{-1}$ in the second term on the left-hand side of \eqref{id1} and then using
\EQ{
\frac{z^{-2}}{x_1^{-1}-z^{-1}}=-\frac1{x_1-z}-\frac1{z}\ .
}

The second identity involves the discontinuity of the integral $\Psi(x_1,x_2)$ across the 
cut $\text x^{(n)}_+$; for $x_1\in \text x^{(n)}_+$:
\EQ{
\Psi(e^\epsilon x_1,x_2)-\Psi(e^{-\epsilon}x_1,x_2)
=-i\log\frac{x_2-z(e^{-\epsilon}x_1)}{x_2-\frac1{z(e^{-\epsilon}x_1)}}\ ,
\label{se1}
}
where $\epsilon$ is an infinitesimally small positive real number.
Here, $z(x)$ is the solution of 
\EQ{
x+\frac1 x+\xi+\frac1\xi=q^{2n}\left(z+\frac1z+\xi+\frac1\xi\right)\ ,
}
with $|z(x)|<1$. This follows from integrating by parts in \eqref{gwe}. When the cut $\text x^{(n)}_+$ is crossed a pair of poles at $z=z(e^{-\epsilon}x)$ and $z(e^{-\epsilon}x)^{-1}$ from the $q$-gamma functions cross the unit circle $|z|=1$ and the difference on the left hand side of \eqref{se1} picks out the residues to give the right-hand side.

The third identity states that for $|x_1^\pm|>1$ and $|x_2|>1$,
\EQ{
\Psi\Big(\frac1{x_1^-},x_2\Big)-\Psi\Big(\frac1{x_1^+},x_2\Big)
=-i\log\Big[\frac{x_2-\frac1{x_1^+}}{x_2}\cdot\frac{x_2-\frac1{x_1^-}}{x_2+\xi}\Big]
\label{id2}
}
which is equivalent to (8.4) of \cite{Arutyunov:2009kf} in the string limit when $\xi\to0$.
The identity relies on writing each function above as an integral using 
\eqref{gwe}. This gives the right-hand side as 
\EQ{
i\oint_{|z|=1}\frac{dz}{2\pi i}\frac1{z-x_2}\log\left[\frac{\Gamma_{q^2}(1+iu(x_1)-iu(z)+\frac12)}{
\Gamma_{q^2}(1-iu(x_1)+iu(z)-\frac12)}\frac{\Gamma_{q^2}(1-iu(x_1)+iu(z)+\frac12)}{
\Gamma_{q^2}(1+iu(x_1)-iu(z)-\frac12)}\right]\ .
}
Then we use the definition of the $q$-gamma function to write this as \eqref{apa}
\EQ{
&i\oint_{|z|=1}\frac{dz}{2\pi i}\frac1{z-x_2}\log\left[\frac{1-q^{2i(u(x_1^-)-u(z))}}{1-q^2}\cdot
\frac{1-q^{2i(u(z)-u(x_1^+))}}{1-q^2}\right]\\
&=i\oint_{|z|=1}\frac{dz}{2\pi i}\frac1{z-x_2}log\left[
\Big(1-\frac{z+\frac1z+\xi+\frac1\xi}{x_1^-+\frac1{x_1^-}+\xi+\frac1\xi}\Big)
\Big(1-\frac{x_1^++\frac1{x_1^+}+\xi+\frac1\xi}{z+\frac1z+\xi+\frac1\xi}\Big)\right]\ .
}
Note that the $1-q^2$ pieces do not contribute.
Integrating by parts and then picking up the poles at $z=0$, $-\xi$, $\frac1{x_1^+}$ and $\frac1{x_1^-}$ gives the result.


\begin{thebibliography}{99}

{\small


%\cite{Beisert:2010jr}
\bibitem{Beisert:2010jr}
  N.~Beisert {\it et al.},
%  ``Review of AdS/CFT Integrability: An Overview,''
  arXiv:1012.3982 [hep-th].
  %%CITATION = ARXIV:1012.3982;%%
%

 %\cite{Beisert:2008tw}
\bibitem{Beisert:2008tw}
  N.~Beisert and P.~Koroteev,
  %``Quantum Deformations of the One-Dimensional Hubbard Model,''
  J.\ Phys.\ A  {\bf 41} (2008) 255204
  [arXiv:0802.0777 [hep-th]].
  %%CITATION = JPAGB,A41,255204;%%

%\cite{arXiv:1002.1097}
\bibitem{arXiv:1002.1097}
  N.~Beisert,
  %``The Classical Trigonometric r-Matrix for the Quantum-Deformed Hubbard Chain,''
  J.\ Phys.\ A\ {\bf 44} (2011) 265202
  [arXiv:1002.1097 [math-ph]].
  %%CITATION = JPAGB,A44,265202;%%
  
 %\cite{Hoare:2011fj}
\bibitem{Hoare:2011fj}
  B.~Hoare and A.~A.~Tseytlin, Nucl.\ Phys.\  {\bf B851} (2011)  161 %``Towards the quantum S-matrix of the Pohlmeyer reduced version of $AdS_5
  %\times S5$ superstring theory,''
  [arXiv:1104.2423 [hep-th]].
  %%CITATION = ARXIV:1104.2423;%%


%\cite{Hoare:2011nd}
\bibitem{Hoare:2011nd}
  B.~Hoare, T.~J.~Hollowood and J.~L.~Miramontes,
  %``A Relativistic Relative of the Magnon S-Matrix,''
  arXiv:1107.0628 [hep-th].
  %%CITATION = ARXIV:1107.0628;%%
  
%\cite{Grigoriev:2007bu}
\bibitem{Grigoriev:2007bu}
  M.~Grigoriev, A.~A.~Tseytlin,
  %``Pohlmeyer reduction of AdS(5) x S**5 superstring sigma model,''
  Nucl.\ Phys.\  {\bf B800 } (2008)  450
  [arXiv:0711.0155 [hep-th]].

%\cite{Mikhailov:2007xr}
\bibitem{Mikhailov:2007xr}
  A.~Mikhailov, S.~Schafer-Nameki,
  %``Sine-Gordon-like action for the Superstring in AdS(5) x S**5,''
  JHEP {\bf 0805 } (2008)  075
  [arXiv:0711.0195 [hep-th]].

%\cite{Grigoriev:2008jq}
\bibitem{Grigoriev:2008jq}
  M.~Grigoriev, A.~A.~Tseytlin,
  %``On reduced models for superstrings on AdS(n) x S**n,''
  Int.\ J.\ Mod.\ Phys.\  {\bf A23 } (2008)  2107
  [arXiv:0806.2623 [hep-th]].

%\cite{Miramontes:2008wt}
\bibitem{Miramontes:2008wt}
  J.~L.~Miramontes,
  %``Pohlmeyer reduction revisited,''
  JHEP {\bf 0810 } (2008)  087
  [arXiv:0808.3365 [hep-th]]. 
  

%\cite{Hollowood:2009tw}
\bibitem{Hollowood:2009tw}
  T.~J.~Hollowood and J.~L.~Miramontes,
  %``Magnons, their Solitonic Avatars and the Pohlmeyer Reduction,''
  JHEP {\bf 0904} (2009) 060
  [arXiv:0902.2405 [hep-th]].
  %%CITATION = JHEPA,0904,060;%%
%\cite{Hollowood:1992sy}
 
 %\cite{Hollowood:2009sc}
\bibitem{Hollowood:2009sc}
  T.~J.~Hollowood and J.~L.~Miramontes,
  %``A New and Elementary CP**n Dyonic Magnon,''
  JHEP {\bf 0908} (2009) 109
  [arXiv:0905.2534 [hep-th]].
  %%CITATION = JHEPA,0908,109;%% 

%\cite{Roiban:2009vh}
\bibitem{Roiban:2009vh}
  R.~Roiban and A.~A.~Tseytlin,
  %``UV finiteness of Pohlmeyer-reduced form of the AdS_5xS^5 superstring
  %theory,''
  JHEP {\bf 0904} (2009) 078
  [arXiv:0902.2489 [hep-th]].
  %%CITATION = JHEPA,0904,078;%%

 %\cite{Hoare:2009rq}
\bibitem{Hoare:2009rq}
  B.~Hoare, Y.~Iwashita and A.~A.~Tseytlin,
  %``Pohlmeyer-reduced form of string theory in AdS_5 x S^5: semiclassical
  %expansion,''
  J.\ Phys.\ A  {\bf 42} (2009) 375204
  [arXiv:0906.3800 [hep-th]].
  %%CITATION = JPAGB,A42,375204;%%
  
  %\cite{Hoare:2009fs}
\bibitem{Hoare:2009fs}
  B.~Hoare and A.~A.~Tseytlin,
  %``Tree-level S-matrix of Pohlmeyer reduced form of AdS(5) x S**5 superstring
  %theory,''
  JHEP {\bf 1002} (2010) 094
  [arXiv:0912.2958 [hep-th]].
  %%CITATION = JHEPA,1002,094;%%

 %\cite{Iwashita:2010tg}
\bibitem{Iwashita:2010tg}
  Y.~Iwashita,
  %``One-loop corrections to AdS_5 x S^5 superstring partition function via
  %Pohlmeyer reduction,''
  J.\ Phys.\ A  {\bf 43} (2010) 345403
  [arXiv:1005.4386 [hep-th]].
  %%CITATION = JPAGB,A43,345403;%%

%\cite{Hollowood:2010rv}
\bibitem{Hollowood:2010rv}
  T.~J.~Hollowood and J.~L.~Miramontes,
  %``The Relativistic Avatars of Giant Magnons and their S-Matrix,''
  JHEP {\bf 1010} (2010) 012
  [arXiv:1006.3667 [hep-th]].
  %%CITATION = JHEPA,1010,012;%%

%\cite{Hoare:2010fb}
\bibitem{Hoare:2010fb}
  B.~Hoare and A.~A.~Tseytlin,
  %``On the perturbative S-matrix of generalized sine-Gordon models,''
  JHEP {\bf 1011} (2010) 111
  [arXiv:1008.4914 [hep-th]].
  %%CITATION = JHEPA,1011,111;%%

%\cite{Hollowood:2010dt}
\bibitem{Hollowood:2010dt}
  T.~J.~Hollowood and J.~L.~Miramontes,
  %``Classical and Quantum Solitons in the Symmetric Space Sine-Gordon
  %Theories,''
  JHEP {\bf 1104} (2011) 119
  [arXiv:1012.0716 [hep-th]].
  %%CITATION = JHEPA,1104,119;%%

%\cite{arXiv:1012.4713}
\bibitem{arXiv:1012.4713}
  D.~M.~Schmidtt,
  %``Supersymmetry Flows, Semi-Symmetric Space Sine-Gordon Models And The Pohlmeyer Reduction,''
  JHEP\ {\bf 1103} (2011) 021
  [arXiv:1012.4713 [hep-th]].
  %%CITATION = JHEPA,1103,021;%%

%\cite{Hollowood:2011fm}
\bibitem{Hollowood:2011fm}
  T.~J.~Hollowood and J.~L.~Miramontes,
  %``The Semi-Classical Spectrum of Solitons and Giant Magnons,''
  JHEP {\bf 1105} (2011) 062
  [arXiv:1103.3148 [hep-th]].
  %%CITATION = JHEPA,1105,062;%%
  
%\cite{Goykhman:2011mq}
\bibitem{Goykhman:2011mq}
  M.~Goykhman and E.~Ivanov,
  %``Worldsheet Supersymmetry of Pohlmeyer-Reduced AdS_n x S^n Superstrings,''
  JHEP {\bf 1109} (2011) 078
  [arXiv:1104.0706 [hep-th]].
  %%CITATION = JHEPA,1109,078;%%

%\cite{Hollowood:2011fq}
\bibitem{Hollowood:2011fq}
  T.~J.~Hollowood, and J.~L.~Miramontes,
  %``The $AdS_5 x S_5$ Semi-Symmetric Space Sine-Gordon Theory,''
  JHEP {\bf 1105 } (2011)  136.
  [arXiv:1104.2429 [hep-th]].
   
 %\cite{arXiv:1106.4796}
\bibitem{arXiv:1106.4796}
  D.~M.~Schmidtt,
  %``Integrability vs Supersymmetry: Poisson Structures of The Pohlmeyer Reduction,''
  JHEP\ {\bf 1111} (2011) 067
  [arXiv:1106.4796 [hep-th]].
  %%CITATION = JHEPA,1111,067;%% 

 %\cite{Iwashita:2011ha}
\bibitem{Iwashita:2011ha}
  Y.~Iwashita, R.~Roiban and A.~A.~Tseytlin,
  %``Two-loop corrections to partition function of Pohlmeyer-reduced theory for
  %AdS_5 x S^5 superstring,''
  arXiv:1109.5361 [hep-th].
  %%CITATION = ARXIV:1109.5361;%%
  
 %\cite{Arutyunov:2004vx}
\bibitem{Arutyunov:2004vx}
  G.~Arutyunov, S.~Frolov and M.~Staudacher,
  %``Bethe ansatz for quantum strings,''
  JHEP {\bf 0410} (2004) 016
  [arXiv:hep-th/0406256].
  %%CITATION = JHEPA,0410,016;%% 

%\cite{Beisert:2005fw}
\bibitem{Beisert:2005fw}
  N.~Beisert, M.~Staudacher,
  %``Long-range psu(2,2|4) Bethe Ansatze for gauge theory and strings,''
  Nucl.\ Phys.\  {\bf B727 } (2005)  1-62.
  [hep-th/0504190].


%\cite{Beisert:2005cw}
\bibitem{Beisert:2005cw}
  N.~Beisert and A.~A.~Tseytlin,
  %``On quantum corrections to spinning strings and Bethe equations,''
  Phys.\ Lett.\ B {\bf 629}, 102 (2005)
  [hep-th/0509084].
  %%CITATION = HEP-TH/0509084;%%


%\cite{Janik:2006dc}
\bibitem{Janik:2006dc}
  R.~A.~Janik,
  %``The AdS(5) x S**5 superstring world-sheet S-matrix and crossing symmetry,''
  Phys.\ Rev.\  D {\bf 73} (2006) 086006
  [arXiv:hep-th/0603038].
  %%CITATION = PHRVA,D73,086006;%%


%\cite{Hernandez:2006tk}
\bibitem{Hernandez:2006tk}
  R.~Hernandez and E.~Lopez,
  %``Quantum corrections to the string Bethe ansatz,''
  JHEP {\bf 0607}, 004 (2006)
  [hep-th/0603204].
  %%CITATION = HEP-TH/0603204;%%




%\cite{Arutyunov:2006iu}
\bibitem{Arutyunov:2006iu}
  G.~Arutyunov, S.~Frolov,
  %``On AdS(5) x S**5 String S-matrix,''
  Phys.\ Lett.\  {\bf B639 } (2006)  378-382.
  [hep-th/0604043].


%\cite{Freyhult:2006vr}
\bibitem{Freyhult:2006vr}
  L.~Freyhult and C.~Kristjansen,
  %``A Universality test of the quantum string Bethe ansatz,''
  Phys.\ Lett.\ B {\bf 638}, 258 (2006)
  [hep-th/0604069].
  %%CITATION = HEP-TH/0604069;%%



%\cite{Beisert:2006ib}
\bibitem{Beisert:2006ib}
  N.~Beisert, R.~Hernandez and E.~Lopez,
  %``A Crossing-symmetric phase for AdS(5) x S**5 strings,''
  JHEP {\bf 0611} (2006) 070
  [arXiv:hep-th/0609044].
  %%CITATION = JHEPA,0611,070;%%

%\cite{Beisert:2006qh}
\bibitem{Beisert:2006qh}
  N.~Beisert,
  %``The Analytic Bethe Ansatz for a Chain with Centrally Extended su(2|2) Symmetry,''
  J.\ Stat.\ Mech.\  {\bf 0701 } (2007)  P01017.
  [nlin/0610017 [nlin.SI]].
  
%\cite{Beisert:2006ez}
\bibitem{Beisert:2006ez}
  N.~Beisert, B.~Eden and M.~Staudacher,
  %``Transcendentality and Crossing,''
  J.\ Stat.\ Mech.\  {\bf 0701}, P01021 (2007)
  [arXiv:hep-th/0610251].
  %%CITATION = JSTAT,0701,P01021;%%

%\cite{Arutyunov:2006yd}
\bibitem{Arutyunov:2006yd}
  G.~Arutyunov, S.~Frolov, M.~Zamaklar,
  %``The Zamolodchikov-Faddeev algebra for AdS(5) x S**5 superstring,''
  JHEP {\bf 0704 } (2007)  002.
  [hep-th/0612229].
  
 
%\cite{Kostov:2007kx}
\bibitem{Kostov:2007kx}
  I.~Kostov, D.~Serban and D.~Volin,
  %``Strong coupling limit of Bethe ansatz equations,''
  Nucl.\ Phys.\  B {\bf 789}, 413 (2008)
  [arXiv:hep-th/0703031].
  %%CITATION = NUPHA,B789,413;%%

%\cite{Dorey:2007xn}
\bibitem{Dorey:2007xn}
  N.~Dorey, D.~M.~Hofman and J.~M.~Maldacena,
  %``On the Singularities of the Magnon S-matrix,''
  Phys.\ Rev.\  D {\bf 76} (2007) 025011
  [arXiv:hep-th/0703104].
  %%CITATION = PHRVA,D76,025011;%%
%\cite{Gromov:2007cd}
\bibitem{Gromov:2007cd}
  N.~Gromov and P.~Vieira,
  %``Constructing the AdS/CFT dressing factor,''
  Nucl.\ Phys.\ B {\bf 790}, 72 (2008)
  [hep-th/0703266].
  %%CITATION = HEP-TH/0703266;%%





%\cite{Arutyunov:2009kf}
\bibitem{Arutyunov:2009kf}
  G.~Arutyunov and S.~Frolov,
  %``The Dressing Factor and Crossing Equations,''
  J.\ Phys.\ A  {\bf 42} (2009) 425401
  [arXiv:0904.4575 [hep-th]].
  %%CITATION = JPAGB,A42,425401;%%

%\cite{Volin:2009uv}
\bibitem{Volin:2009uv}
  D.~Volin,
  %``Minimal solution of the AdS/CFT crossing equation,''
  J.\ Phys.\ A  {\bf 42} (2009) 372001
  [arXiv:0904.4929 [hep-th]].
  %%CITATION = JPAGB,A42,372001;%%

%\cite{Kruczenski:2009kc}
\bibitem{Kruczenski:2009kc}
  M.~Kruczenski and A.~Tirziu,
  %``On the dressing phase in the SL(2) Bethe Ansatz,''
  Phys.\ Rev.\ D {\bf 80}, 086002 (2009)
  [arXiv:0907.4118 [hep-th]].
  %%CITATION = ARXIV:0907.4118;%%




%\cite{Vieira:2010kb}
\bibitem{Vieira:2010kb}
  P.~Vieira and D.~Volin,
  %``Review of AdS/CFT Integrability, Chapter III.3: The Dressing factor,''
  arXiv:1012.3992 [hep-th].
  %%CITATION = ARXIV:1012.3992;%%


%\cite{arXiv:0710.1568}
\bibitem{arXiv:0710.1568}
  G.~Arutyunov and S.~Frolov,
  %``On String S-matrix, Bound States and TBA,''
  JHEP\ {\bf 0712} (2007) 024
  [arXiv:0710.1568 [hep-th]].
  %%CITATION = JHEPA,0712,024;%% 

%\cite{arXiv:0901.1417}
\bibitem{arXiv:0901.1417}
  G.~Arutyunov and S.~Frolov,
  %``String hypothesis for the AdS(5) x S**5 mirror,''
  JHEP\ {\bf 0903} (2009) 152
  [arXiv:0901.1417 [hep-th]].
  %%CITATION = JHEPA,0903,152;%% 

%\cite{Bombardelli:2009ns}
\bibitem{Bombardelli:2009ns}
  D.~Bombardelli, D.~Fioravanti and R.~Tateo,
  %``Thermodynamic Bethe Ansatz for planar AdS/CFT: a proposal,''
  J.\ Phys.\ A  {\bf 42} (2009) 375401
  [arXiv:0902.3930 [hep-th]].
  %%CITATION = JPAGB,A42,375401;%%

%\cite{arXiv:0902.4458}
\bibitem{arXiv:0902.4458}
  N.~Gromov, V.~Kazakov, A.~Kozak and P.~Vieira,
  %``Exact Spectrum of Anomalous Dimensions of Planar N = 4 Supersymmetric Yang-Mills Theory: TBA and excited states,''
  Lett.\ Math.\ Phys.\ \ {\bf 91} (2010) 265
  [arXiv:0902.4458 [hep-th]].
  %%CITATION = LMPHD,91,265;%%
  
%\cite{arXiv:0907.2647}
\bibitem{arXiv:0907.2647}
  G.~Arutyunov and S.~Frolov,
  %``Simplified TBA equations of the AdS(5) x S**5 mirror model,''
  JHEP\ {\bf 0911} (2009) 019
  [arXiv:0907.2647 [hep-th]].
  %%CITATION = JHEPA,0911,019;%% 
 
%\cite{arXiv:1012.3995}
\bibitem{arXiv:1012.3995}
  Z.~Bajnok,
  %``Review of AdS/CFT Integrability, Chapter III.6: Thermodynamic Bethe Ansatz,''
  arXiv:1012.3995 [hep-th].
  %%CITATION = ARXIV:1012.3995;%%  
  
%\cite{deLeeuw:2011jr}
\bibitem{deLeeuw:2011jr}
  M.~de Leeuw, V.~Regelskis, T.~Matsumoto,
  %``The Bound State S-matrix of the Deformed Hubbard Chain,''
  arXiv:1109.1410 [math-ph].
  
%\cite{Beisert:2011wq}
\bibitem{Beisert:2011wq}
  N.~Beisert, W.~Galleas and T.~Matsumoto,
  %``A Quantum Affine Algebra for the Deformed Hubbard Chain,''
  arXiv:1102.5700 [math-ph].
  %%CITATION = ARXIV:1102.5700;%%

%\cite{Miramontes:1999gd}
\bibitem{Miramontes:1999gd}
  J.~L.~Miramontes,
  %``Hermitian analyticity versus real analyticity in two-dimensional factorized S matrix theories,''
  Phys.\ Lett.\ B {\bf 455} (1999) 231
  [hep-th/9901145].
  %%CITATION = HEP-TH/9901145;%%  

%\cite{Dorey:1996gd}
\bibitem{Dorey:1996gd}
  P.~Dorey,
  ``Exact S matrices,''
   [hep-th/9810026].  

%\cite{hep-th/0511082}
\bibitem{hep-th/0511082}
  N.~Beisert,
  %``The SU(2|2) dynamic S-matrix,''
  Adv.\ Theor.\ Math.\ Phys.\ \ {\bf 12} (2008) 945
  [hep-th/0511082].
  %%CITATION = 00203,12,945;%%  
  
%\cite{hep-th/0606214}
\bibitem{hep-th/0606214}
  N.~Beisert,
  %``On the scattering phase for AdS(5) x S**5 strings,''
  Mod.\ Phys.\ Lett.\ A\ {\bf 22} (2007) 415
  [hep-th/0606214].
  %%CITATION = MPLAE,A22,415;%%  
  
%\cite{Arutyunov:2009ga}
\bibitem{Arutyunov:2009ga}
  G.~Arutyunov, S.~Frolov,
  %``Foundations of the AdS_5 x S^5 Superstring. Part I,''
  J.\ Phys.\ A {\bf A42}, 254003 (2009).
  [arXiv:0901.4937 [hep-th]].
  
 

}

\end{thebibliography}
\end{document}